\documentclass[prb,twocolumn,nobibnotes,groupedaddress,superscriptaddress,floatfix,a4paper]{revtex4-2} 

\usepackage{amsmath}
\usepackage{amssymb}
\usepackage{mathtools}
\usepackage{upgreek}
\usepackage{braket}
\usepackage[normalem]{ulem}
\usepackage{graphicx}
\usepackage{xcolor} 

\usepackage{booktabs}
\usepackage{multirow}
\usepackage{array} 

\usepackage{enumerate}
\usepackage{algorithm}
\usepackage{algpseudocode}

\usepackage[utf8]{inputenc}
\usepackage[colorlinks=true,linkcolor=blue,citecolor=blue,urlcolor=blue]{hyperref}

\newcommand{\dan}[1]{{\color{blue}$^\text{Dan}$(#1)}}

\DeclareMathOperator*{\argminA}{arg\,min}

\begin{document}
\title{Resolving molecular biexcited states using classical leading-order triples corrections to quantum equation-of-motion UCCSD}
\thanks{This manuscript has been authored by UT-Battelle, LLC, under Contract No.~DE-AC0500OR22725 with the U.S.~Department of Energy. The United States Government retains and the publisher, by accepting the article for publication, acknowledges that the United States Government retains a non-exclusive, paid-up, irrevocable, world-wide license to publish or reproduce the published form of this manuscript, or allow others to do so, for the United States Government purposes. The Department of Energy will provide public access to these results of federally sponsored research in accordance with the DOE Public Access Plan.}

\author{Zachary W. Windom}
\email{windomzw@ornl.gov}
\affiliation{Quantum Information Science Section, Oak Ridge National Laboratory, Oak Ridge, TN 37831, USA}

\author{Prince F. Kwao}
\affiliation{Department of Chemistry, University of North Dakota, Grand Forks, ND 58202, USA}

\author{Luke Bertels}
\affiliation{Quantum Information Science Section, Oak Ridge National Laboratory, Oak Ridge, TN 37831, USA}

\author{Daniel Claudino}
\email{claudinodc@ornl.gov}
\affiliation{Quantum Information Science Section, Oak Ridge National Laboratory, Oak Ridge, TN 37831, USA}

\author{Ayush Asthana}
\email{ayush.asthana@und.edu}
\affiliation{Department of Chemistry, University of North Dakota, Grand Forks, ND 58202, USA}

\author{Rodney J. Bartlett}
\affiliation{Quantum Theory Project, University of Florida, Gainesville, FL 32611, USA}

\begin{abstract}
Excited electronic states are difficult to accurately and tractably model, particularly those exhibiting multiexcitonic character. Scalable single-reference excited state methods typically fail to describe such states as they suffer from the defects of the mean-field approximation and/or neglect higher-rank excitation operators. To address this, we derive the [T] perturbative correction to the quantum self-consistent equation-of-motion unitary coupled cluster singles and doubles (q-sc-EOM-UCCSD) method. The method captures triple excitation effects through a classical step of second-order in many-body perturbation theory following a q-sc-EOM-UCCSD calculation on a quantum computer.  The perturbative correction developed has a $\mathcal{O}(N^7)$ classical computational time scaling.  We incorporate [T] into a hybrid compute strategy wherein eigenvectors of the effective Hamiltonian from the hybrid quantum-classical q-sc-EOM-UCCSD algorithm are relayed to a classical computer which performs the postprocessing required for the [T] correction. The benefits of [T] are quantified in a benchmark covering a variety of electronically excited states of the isoelectronic CH$^+$ and BH molecules. We find that the addition of the [T] correction to both trotterized and full operator variants of q-sc-EOM-CCSD can offer dramatic improvements over baseline q-sc-EOM-UCCSD. Our assessment clearly demonstrates the importance of considering higher-rank excitation operators in q-sc-EOM-UCCSD, quantifies the overall success of the [T] correction, and discusses some of its limitations.

\end{abstract}

\maketitle

\section{Introduction}

Biexcited states facilitate a variety of biological processes and are integral in several technological applications. Resolving these two-photon mediated, optically ``dark" states require nonlinear spectroscopies as this type of excited state does not necessarily obey traditional selection rules. Nevertheless, a multitude of physical phenomena are mediated by biexcited states; for example,  up- and down-conversion of energy through processes found in singlet-fission,\cite{kim2018coupled} multiexciton generation in quantum dots,\cite{nozik2008multiple} as well as light-harvesting devices like solar cells.\cite{davis2015multiple} These states are also important in enzymatic reactions facilitating photosynthesis and eyesight.\cite{accomasso2022ultrafast,polivka2004ultrafast,mirkovic2017light} The lifetime of such states are anomalously long (up to 1 ms), leading to long coherence times. Consequently, these states naturally support qubit and quantum memory technologies that are currently of interest in the quantum information science community.\cite{scholes2025quantum} With this in mind, the current work intends to improve the \textit{ab initio} resolution of doubly excited states.

Electronic structure method development targeting biexcited states is an active area of research,\cite{boguslawski2018targeting,lee2019excited,loos2019reference,kossoski2021excited,ravi2023excited,rishi2023dark,ravi2022intermediate}
suggesting that a robust theoretical treatment for these states is still lacking. This is quite unlike the techniques designed to capture states dominated by single excitations. For example, the equation of motion coupled cluster singles and doubles (EOM-CCSD) approximation~\cite{koch1990excitation,geertsen1989equation} has been shown to predict singles dominated excitations to within 0.15-0.3 eV\cite{stanton1993equation}---which can be further reduced to $\approx$0.01 eV if perturbative triples are added.\cite{watson2013benchmarking} Despite this success, such techniques incur errors that can exceed 3.95 eV for excited states that are dominated by two-particle excitation effects.\cite{watts1994inclusion,watts1995economical,rishi2017excited,kowalski2001active} 
 
This extreme  difference in performance is, at this point, well understood in traditional EOM theory. Singles dominated states rely on adequate description of two-particle excitations to provide the requisite amplitude relaxation effects. The situation is similar for the description of doubles dominated excited states, which require an accounting of triple excitations.\cite{watts1996iterative} This alludes to the fact that such errors can, in principle, be ameliorated by increasing the cluster operator rank,\cite{christiansen1995response,bartlett2007coupled,windoma2024t2,windom2024ultimate}  which can mean incorporating triples and/or quadruples amplitudes (via EOM-CCSDT\cite{kucharski2001coupled} and/or EOM-CCSDTQ.\cite{kallay2001higher,hirata2000high}) With this rule of thumb in mind, conservative prediction of doubly excited  electronic states should require an adequate accounting of triple excitations---at least---for predictive accuracy. However, adding higher-rank cluster operators assumes the resource requirements and computational complexity tradeoffs---which scale polynomially with system size---can be afforded.


To this point, one of the primary motivations guiding the development of quantum computing pursues the prospect of an exponential  speedup over conventional approaches using classical computers. Algorithms such as the quantum phase estimation (QPE)\cite{kitaev1997quantum,nielsen2010quantum,russo2021evaluating,bauman2020toward,sugisaki2021quantum,santagati2018witnessing,o2019quantum} and others\cite{motta2020determining,parrish2019quantum,sugisaki2022adiabatic} are formally exact and are expected to require dramatically less computational time and memory compared to the conventional classical approach. However, algorithms of this type require an enormous number of physical qubits, with extremely low error rates and often error correction protocols characteristic of the fault-tolerant era of quantum computing. Unfortunately, these resource requirements are untenable in the expected hardware that will be available in the near term, which is expected to have a limited number of noisy, error-prone physical qubits. 

However, recent near-term algorithms based on the circuit-depth friendly variational quantum eigensolver(VQE) have been proposed for ground state \cite{peruzzo2014variational,kandala2017hardware,romero2018strategies,ryabinkin2018qubit,grimsley2019adaptive,tang2021qubit} and excited states\cite{chan2021molecular,greenediniz2021generalized,ryabinkin2018constrained,parrish2019mcvqe}. More recent multi-state adaptive formulations include MORE-ADAPT-VQE and TEPID-ADAPT-VQE, which target several low-lying states within a common optimization.\cite{grimsley2025challenging,sambasivam2025tepid,saroni2026tepid} Several algorithms proposed for molecular excited states for applications on near-term quantum hardware can broadly be categorized into two types: (i) diagonalization based strategies, or (ii) circuit optimisation-based approaches. The latter involves successive state preparation and circuit optimization for each excited state by biasing the cost function with respect to a single or multiple excited states; examples include subspace-search VQE (SS-VQE),\cite{nakanishi2019subspace} variational quantum deflation (VQD),\cite{chan2021molecular,higgott2019variational} the folded spectrum method,\cite{peruzzo2014variational} and the orthogonal state reduction variational eigensolver (OSRVE).\cite{xie2022orthogonal} However, quantum resource requirements of these approaches increase - particularly in relation to the circuit depth---upon solution of successively higher energy solutions. Diagonalization strategies, on the other hand, use a classical computer to diagonalize the Hamiltonian projected on a subspace to estimate multiple solutions simultaneously. Examples include quantum Krylov subspace expansion,\cite{motta2020determining,tkachenko2024quantum} quantum subspace expansion,\cite{mcclean2017hybrid,colless2018computation,takeshita2020increasing,mcclean2020decoding} and the quantum equation of motion (qEOM).\cite{ollitrault2020quantum} The qEOM method in particular appears interesting as it aspires to be a quantum and unitary analog to the conventional EOM approach for excited states.


However, it is now recognized that the original qEOM procedure\cite{ollitrault2020quantum} does not satisfy exact conditions; specifically,  the ``killer condition"---equivalently known as the Vacuum Annihilation Condition (VAC)---which asserts that the ground state reference cannot be deexcited. A consequence of this theoretical flaw manifests as nonphysical errors creeping into ionization potential (IPs) and electron affinity (EAs) calculations. Fortunately, the solution to this problem has previously been established and involves a self-consistent transformation of the qEOM operator basis. Incorporating this procedure into the qEOM procedure, Asthana et al. derived the quantum self-consistent EOM (q-sc-EOM) method, where demonstrations reveal not only fidelity improvements but also significant reductions in measurement requirements and resilience to sampling and other errors.\cite{asthana2023quantum,kwao2026generalized} However, the ability of such low-rank techniques to accurately resolve biexcited states using only singles and doubles excitation operators remains suspect. Complementary work has combined ADAPT-VQE~\cite{grimsley2019adaptive} or local unitary cluster jastrow (LUCJ)~\cite{Motta2023} reference-state preparation with q-sc-EOM and demonstrated resource-reduced, error-mitigated excited-state calculations on quantum hardware.\cite{sundar2026molecular}



Various flavors of perturbation theory have been designed over the years as a way to reliably improve prediction fidelity that while also largely circumnavigating existing resource limitations.\cite{bartlett1990non,watts1995economical,watts1996iterative,stanton1996simple,patra2025energy} Note that such techniques have been established in relation to both conventional and unitary coupled cluster theories,\cite{windom2024new,windom2025toward,windom2024attractive,windom2026analysis,zhao2024enhancing} with recent efforts focusing on engineering classically tractable unitary coupled cluster (UCC) approximations.\cite{bartlett1988expectation,bartlett1989some,bartlett1989alternative,taube2006new} One notable example invokes the properties of Bernoulli polynomials to simplify the structure of the Baker-Campbell-Hausdorff (BCH) expansion.\cite{liu2021unitary,hodecker2020third,kutzelnigg1983quantum,liu2018unitary} The resulting series has been manipulated in various ways, leading to an assortment of approximations for both the ground and excited state that are informed by an underlying perturbation theory. This is perhaps best illustrated in the recent work of Liu et al\cite{liu2022quadratic} where a triples correction is defined in relation to the Lowdin's partitioning approach.\cite{lowdin1962studies} Also related to the current work, Zheng et al. recently developed a perturbative triples correction to q-sc-EOM and QSE that is performed entirely on quantum hardware and requires the construction of higher-order reduced density matrices (RDMs).\cite{zheng2024quantum} Caution is warranted when using RDM-based perturbation theories, however, as such approximations can introduce non-cumulant terms into the energy which can sacrifice size-intensivity; further discussion of this topic is deferred to the Conclusion.


The current work expands upon these prior efforts by deriving a perturbative triples correction to q-sc-EOM-UCCSD, which we call [T]. To do this, we leverage a hybrid compute strategy that partitions the workload between the quantum device and classical computer with the objective of increasing prediction fidelity at no additional quantum resource requirement. Specifically, the quantum device is responsible for computing ground and excited state q-sc-EOM-UCCSD amplitudes. The amplitudes, alongside the Hamiltonian, are then leveraged by a classical computer to compute the leading-order triples correction, [T], to q-sc-EOM-UCCSD. This correction is rigorously size-intensive, and requires a non-iterative $\mathcal{O}(N^7)$ classical algorithm without necessitating any quantum resources beyond baseline q-sc-EOM. We interrogate the utility of [T] by benchmarking its performance for various singly- and doubly-excited states. Our analysis considers the BH and CH$^+$ molecules at different geometries, and intends to quantify the limitations of q-sc-EOM-UCCSD and survey the extent upon which the [T] correction can improve baseline predictions.

\section{Theory}
\label{sec:theory}

The electronic correlation energy is defined with respect to the normal-ordered Hamiltonian $H_N$
\begin{equation}
\begin{split}
\label{eq:ham}
H_N&=H-\braket{0|H|0}\\
&=\underbrace{\sum_{p}f_{pp}\{p^{\dagger}p\}   }_{\text{$H_0$}}+ \underbrace{ \frac{1}{4}\sum_{pqrs}\braket{pq||rs}\{p^{\dagger}q^{\dagger}sr\}}_{\text{$W_N$}} \\
& =  H_0 + W_N,
  \end{split}
\end{equation}
where $p^\dagger$ and $q$ are creation and annihilation operators that act on the spin-orbital indexed by $p$/$q$, which are normal-ordered when within $\{\}$. The mean-field reference function $\ket{0}$ is chosen to be the Hartree-Fock determinant. Note that $\braket{pq||rs}$ refers to the set of anti-symmetrized two-electron integrals, and $f_{pp}$ refers to diagonal elements of the Fock matrix. We have further partitioned $H_N$ in Equation \eqref{eq:ham} into an ``easy to solve", zeroth-order component, $H_0$, and a perturbation, $W_N$, for reasons that will become clear in our ensuing derivations. 

The following work assumes an exponential parameterization of the ground state wavefunction of the form
\begin{equation}
    \ket{\psi_0}=e^{\tau}\ket{0},
\end{equation} 
In this case, $\tau = \sum_n^N \tau_n$ represents the anti-Hermitian cluster operator which modulates the excitation/de-excitation of $n$ electrons out of the reference space describing a $N$-electron problem. Similar to conventional coupled cluster (CC) theory, its unitary variant (UCC) similarly becomes exact if the cluster operator accounts for all $n\in [1,\cdots,N]$ excitations/de-excitations. The cluster operator, $\tau_n$, can also be defined in terms of conventional cluster amplitudes (e.g. $t_{ij\cdots}^{ab\cdots}$) via the relation
\begin{equation}
\begin{split}
 \tau_n &= T_n - T_n^{\dagger} \\
    &=\frac{1}{(n!)^2}\sum_{ab\cdots ij\cdots} \left(t_{ij\cdots}^{ab\cdots}\{a^{\dagger}b^{\dagger}\cdots ji\}-t_{ab\cdots}^{ij\cdots}\{i^{\dagger}j^{\dagger}\cdots ba\}\right).
\end{split}
\end{equation}

The current work assumes a ground state ansatz that considers singles and doubles excitation/de-excitation operators, $\tau\equiv \tau_1+\tau_2$, which in turn defines the UCCSD approximation.

\subsection{Background for EOM-UCCSD}
\label{ssec:eom-uccsd}

A common  approach to derive the  conventional EOM equations begins by applying a similarity transformation to the normal-ordered Hamiltonian, defined in Equation \eqref{eq:ham}.\cite{shavitt2009many} This philosophy can similarly be leveraged to construct a unitary analog of conventional EOM: specifically, the ground state UCCSD wavefunction is used to define an ``effective" Hamiltonian, $\mathcal{H}$,  such that
\begin{equation}\label{eq:Hbar}
    \mathcal{H} \equiv e^{-\tau_1 -\tau_2}H_Ne^{\tau_1 +\tau_2}.
\end{equation}
The ground-state UCCSD correlation energy, $\Delta E_0$, can be defined with respect to this operator according to the eigenvalue equation
\begin{equation}\label{eq:groundStateEqn}
    \mathcal{H} \ket{0} = \Delta E_0 \ket{0},
\end{equation} which conveniently has the non-canonical, mean-field reference $\ket{0}$, as an eigenvector and $\Delta E_0=E_0 -\braket{0|H|0}$.

Despite representing a suitable starting point for the development of UCC ground-state theories, extending Equation \eqref{eq:groundStateEqn} for excited electronic states requires ensuring certain theoretical conditions are satisfied; we deal with these considerations now. Subsequent discussion of EOM-UCC theory will be based on a rearranged---but equivalent---form of Equation \eqref{eq:groundStateEqn}, that is used for convenience, expressed as 
\begin{equation}\label{eq:groundStateEqn1}
    \begin{split}
        H_N\ket{\psi_0}=\Delta E_0\ket{\psi_0}.
    \end{split}
\end{equation}

The $k$-th excited-state solution, $\ket{\Psi^k}$, can be generated by application of a linear, anti-Hermitian excitation operator, $\mathcal{R}^k$, to the ground state (UCCSD) wavefunction
\begin{equation}
    \ket{\Psi^k} \equiv \mathcal{R}^k\ket{\psi_0},
\end{equation} where $\mathcal{R}^k$ is equivalently referred to as a state-transfer operator. It is critically important to recognize that there are two properties $\mathcal{R}^k$ must satisfy to ensure the integrity of the theory. The first property emphasizes the importance of $(\mathcal{R}^k)^{\dagger}$ satisfying the vacuum annihilation condition (VAC), particularly when $\mathcal{R}^k$ is inexact. Since the eigenstates of an exact theory are orthogonal, the VAC is automatically guaranteed if the exact ground state and state-transfer operator are used. To this point, satisfaction of the VAC is ensured if the ground state cannot be de-excited\cite{prasad1985some,szekeres2001killer,mertins1996algebraic,weiner1980calculation,hodecker2020unitary} 
\begin{equation}\label{eq:VAC}
    \big(\mathcal{R}^k\big)^{\dagger}\ket{\psi_0} = 0, \forall k
\end{equation} because of the orthogonality condition that enforces $\ket{\psi_0} \perp\ket{\Psi^k}$. Equation \eqref{eq:VAC} further implies that $\big(\mathcal{R}^k\big)^{\dagger} \equiv \ket{\psi_0}\bra{\Psi^k}$. 

Fortunately, satisfaction of the VAC using inexact operators and states can nevertheless be achieved by using ``self-consistent" operators as originally proposed by Mukherjee et al.\cite{prasad1985some,datta1993consistent} To generate these self-consistent operators, an excitation manifold accounting for all possible single and double excitations and de-excitations is constructed where $\{ \mathcal{A}_I\} \cup \{ \mathcal{A}_I^{\dagger}\}$, and $\mathcal{A}_I$ corresponds to any single, $(a^{\dagger}i )$, or double, $(a^{\dagger}b^{\dagger}ji )$, excitation operator. Rotation of this excitation manifold then generates the so-called self-consistent manifold, $\{ \mathcal{B}_I\} \cup \{ \mathcal{B}_I^{\dagger}\}$, where 
\begin{equation}\label{eq:NewExcOp}
    \mathcal{B}_I = e^{\tau}\mathcal{A}_Ie^{-\tau}.
\end{equation} Satisfaction of the VAC can then be facilitated by using this self-consistently transformed operator manifold as the foundation upon which approximate excited state theories can be reliably built.

With this in mind, the state-transfer operator governing electronic excitations, $\{ \mathcal{R}^k \}$,  can be expressed in terms of a linear combination of all possible operators contained within the self-consistent excitation manifold:
\begin{equation}
\begin{split}\label{eq:fullR}
        \mathcal{R}^k = R_1^k + R_2^k,
\end{split}
\end{equation} where $ R_1^k $ and $ R_2^k$ are  single and double excitation operators, defined according to \begin{equation}\label{eq:sc2}
    \begin{split}
        &R_1^k = \sum_{ia} r_i^a \mathcal{B}_i^a - r_a^i\mathcal{B}_a^i, \\
        &R_2^k = \sum_{ijab} r_{ij}^{ab} \mathcal{B}_{ij}^{ab} - r_{ab}^{ij}\mathcal{B}_{ab}^{ij}. \\
    \end{split}
\end{equation} Equation \eqref{eq:sc2} defines the excitation operators $R_1^k$ and $R_2^k$  with respect to  excitation (e.g. $r_i^a$) and de-excitation (e.g. $r_a^i$) amplitudes. It is important to stress that - until this point - EOM-UCC theory is clearly distinct from its conventional analog since  its set of ground and excited state excitation/de-excitation operators do not, in general, commute. This make EOM-UCC theory unique, and  warrants alternative  avenues for its formulation. 

For notational convenience, we drop the excited-state superscript label $k$ in $\mathcal{R}^k$ and tangentially note that Equations \eqref{eq:VAC}, \eqref{eq:NewExcOp}, \eqref{eq:fullR}, and \eqref{eq:sc2} can be used to define another eigenvalue equation describing electronically excited states
\begin{equation}\label{eq:EEeig}
    H_N \mathcal{R}\ket{\psi_0} = \Delta E_k\mathcal{R}\ket{\psi_0},
\end{equation} where $\Delta E_k$ represents electron correlation contributions to the $k$-th excited-state's electronic energy such that $\Delta E_k=E_k -\braket{0|H|0}$.

After inserting Equations \eqref{eq:NewExcOp}, \eqref{eq:fullR}, and \eqref{eq:sc2} into Equation \eqref{eq:EEeig}, manipulating the algebra reveals an interesting relationship in the excited state formalism:

\begin{equation}
\begin{split}
   \underbrace{e^{-\tau}H_Ne^{\tau}}_{\equiv\mathcal{H}}\underbrace{\left(\sum_{a,b,\dots,i,j,\dots}r_{ij\cdots}^{ab\cdots}\{a^{\dagger}b^{\dagger}\cdots ji\}\right)}_{\equiv \mathcal{C}}\ket{0}= \\
   \Delta E_k \left(\sum_{a,b,\dots,i,j,\dots}r_{ij\cdots}^{ab\cdots}\{a^{\dagger}b^{\dagger}\cdots ji\}\right) \ket{0},
\end{split}
\end{equation} which is clearly expressed in terms of the ``effective" Hamiltonian as originally described in Equation \eqref{eq:Hbar} and a new operator, $\mathcal{C}\equiv \left(\sum_{a,b,\dots,i,j,\dots}r_{ij\cdots}^{ab\cdots}\{a^{\dagger}b^{\dagger}\cdots ji\}\right)$. For brevity, the final excited-state eigenvalue expression can be written as
\begin{equation}\label{eq:finalEkeqn}
    \mathcal{H}\mathcal{C}\ket{0}=\Delta E_k \mathcal{C}\ket{0}.
\end{equation}

A similar strategy can be applied to the ground state eigenvalue expression. Specifically, $\mathcal{R}$ is applied to the left of Equation \eqref{eq:groundStateEqn1}, yielding
\begin{equation}\label{eq:reformE0eqn}
    \mathcal{R}H_N\ket{\psi_0}=\Delta E_0\mathcal{R}\ket{\psi_0}.
\end{equation}
Equation \eqref{eq:reformE0eqn} can, in turn, be rewritten using Equations \eqref{eq:NewExcOp}, \eqref{eq:fullR}, and \eqref{eq:sc2} such that
\begin{equation}\label{eq:notfinalE0}
           e^{\tau}\mathcal{C}\underbrace{e^{-\tau}H_Ne^{\tau}}_{\equiv\mathcal{H}}\ket{0} = \Delta E_0e^{\tau}\mathcal{C}e^{-\tau}e^{\tau}\ket{0} ,
\end{equation} where we have both used the definition of $\mathcal{C}$ and note the  presence of $\mathcal{H}$ as in the prior excited-state example. Algebraic rearrangement of Equation \eqref{eq:notfinalE0} reveals the following:
\begin{equation}\label{eq:finalE0eqn}
    \mathcal{C}\mathcal{H}\ket{0} = \Delta E_0 \mathcal{C}\ket{0},
\end{equation}

Now, we can subtract Equation \eqref{eq:finalE0eqn} from Equation \eqref{eq:finalEkeqn} to yield a commutator-based eigenvalue equation which is defined in relation to the energy difference between the ground and excited state 
\begin{equation}
\begin{split}\label{eq:EOMUCCeq}
      \left(\mathcal{H}\mathcal{C} - \mathcal{C}\mathcal{H} \right)\ket{0} &= \left( \Delta E_k-\Delta E_0\right)\mathcal{C}\ket{0} \\  
     [\mathcal{H},\mathcal{C}]\ket{0} &= \omega_k \mathcal{C}\ket{0}.
\end{split}
\end{equation} 
That energy difference corresponds to the excitation energy, $\omega_k = \Delta E_k-\Delta E_0$, which is a size-intensive quantity by design as it is determined by taking the difference between two size-extensive quantities. Also note that the same reference energy, $\braket{0|H|0}$, has been subtracted from both $E_k$ and $E_0$.

Due to the Hermitian symmetry of $\mathcal{H}$, the adjoint of Equation \eqref{eq:EOMUCCeq} can equivalently be used to determine the excitation energy
\begin{equation}
    \bra{0}[\mathcal{C}^{\dagger},\mathcal{H}] = \omega_k \bra{0}\mathcal{C}^{\dagger}.
\end{equation}

Assuming that the corresponding BCH expansion has not been artificially truncated, the EOM-UCC excitation energy is constructed from linked components as its commutator structure involving $\mathcal{C}$ and $\tau$ mediates the removal of disconnected diagrams. This is in direct contrast to the situation found in conventional EOM-CC theory where the left-hand eigenvectors are not duals of the right-hand solutions, which in turn fosters the possibility of internally disconnected components in the left-hand, bra states, specifically. Unlike in conventional EOM-CC theory, the excitation energy found in EOM-UCC theory can be equivalently expressed in two  ways:
\begin{equation}\label{eq:connect}
       \omega_k = \braket{0|[\mathcal{C}^{\dagger},[\mathcal{H},\mathcal{C}]] |0} =\braket{0|[[\mathcal{C}^{\dagger},\mathcal{H}],\mathcal{C}]|0}, 
\end{equation} assuming normalization. The excitation energy depends explicitly on the excitation amplitudes, $\mathcal{C}$, and implicitly depends on the cluster amplitudes for the ground state, $\tau$. Additionally, the equations defining the excitation operator, $\mathcal{C}$,  \textbf{must} be the Hermitian conjugate of the equations defining  $\mathcal{C}^{\dagger}$. This fact is a direct consequence of the underlying Hermitian symmetry which governs $\mathcal{H}$, and can also be used to rationalize the nested commutator structure of Equation \eqref{eq:connect}.

\subsection{Leading-order triples corrections to EOM-UCCSD}

To determine the leading-order triples correction to baseline EOM-UCCSD, we first approximate the full similarity-transformed Hamiltonian of Equation \eqref{eq:Hbar} through fourth-order, such that $\mathcal{H}\approx\mathcal{H}_{\text{UCCSD(4)}}$ where

\begin{equation}
    \mathcal{H}_{\text{UCCSD(4)}} = \mathcal{H}^{[0]}+\mathcal{H}^{[1]}+\mathcal{H}^{[2]}+\mathcal{H}^{[3]} +\mathcal{H}^{[4]}
\end{equation} and

\begin{widetext}
\begin{equation}\label{eq:HbarUCC4}
    \begin{split}
        &\mathcal{H}^{[0]} = H_0, \\
        &\mathcal{H}^{[1]} = W_N + [H_0,\tau_2], \\
        &\mathcal{H}^{[2]} = [H_0,\tau_1] + [W_N,\tau_2]+ \frac{1}{2}[[H_0,\tau_2],\tau_2], \\
        &\mathcal{H}^{[3]} = [W_N,\tau_1] + \frac{1}{2}\bigg([[W_N,\tau_2],\tau_2] + [[H_0,\tau_1],\tau_2] + [[H_0,\tau_2],\tau_1] \bigg) + \frac{1}{3!}[[[H_0,\tau_2],\tau_2],\tau_2],\\
        & \mathcal{H}^{[4]}= \frac{1}{3!}\bigg([[[W_N,\tau_2],\tau_2],\tau_2]+ [[[H_0,\tau_1],\tau_2],\tau_2] +[[[H_0,\tau_2],\tau_1],\tau_2]+[[[H_0,\tau_2],\tau_2],\tau_1]\bigg)+\cdots
    \end{split}
\end{equation} 
\end{widetext}
are order-by-order elements of the fourth-order UCCSD(4) energy functional.\cite{bartlett1989some,bartlett1989alternative,windom2024new} Prior work argues that the full fourth-order energy functional expressed in Equation \eqref{eq:HbarUCC4} can be dramatically simplified by invoking lower-order residual equations, meaning that several terms of the above are effectively zero.\cite{bartlett1989some,bartlett1989alternative,windom2024new,windoma2024t2,windom2024attractive,windom2026analysis,windom2025toward,windom2024ultimate} This is important to point out, as we will rely on this concept to simplify the excited state energy functional in subsequent derivations. 

Our zeroth-order---or equivalently, P-space---approximation is defined with respect to all singly and doubly excited determinants, $\ket{\textbf{h}_1\textbf{h}_2}$. The reference state UCCSD(4) equations are then expressed as 
\begin{equation}
    \begin{split}
        &\braket{\textbf{h}_1|\mathcal{H}_{\text{UCCSD(4)}}|0}= \braket{0|\mathcal{H}_{\text{UCCSD(4)}}|\textbf{h}_1}=0,\\
        &\braket{\textbf{h}_2|\mathcal{H}_{\text{UCCSD(4)}}|0}= \braket{0|\mathcal{H}_{\text{UCCSD(4)}}|\textbf{h}_2} = 0,\\
        &\braket{0|\mathcal{H}_{\text{UCCSD(4)}}|0}=\Delta E_{\text{UCCSD(4)}},
    \end{split}
\end{equation}
where the first two lines are the singles and doubles residual equations and the final line is the expression defining the UCCSD(4) ground state correlation energy. Using the equations derived in Sec. \ref{ssec:eom-uccsd}, we similarly define the q-sc-EOM-UCCSD(4) correlation energy for the excited state. 
We also define this as the zeroth-order part of our problem such that $\mathcal{C}_k\equiv C_1 + C_2 =C_P$ is used within Equation \eqref{eq:connect}:
\begin{equation}
    \omega_{\text{UCCSD(4)}}\braket{0|[C_P^{\dagger},C_P]|0} = \braket{0|[C_P^{\dagger},[\mathcal{H}_{\text{UCCSD(4)}}, C_P]]|0},
\end{equation} where Equation \eqref{eq:HbarUCC4} has been used in place of $\mathcal{H}$ and the overlap is kept on the LHS of the equation for reasons that will become clear shortly. Note that there is a distinction between the meaning of ``order" when defining the effective Hamiltonian of UCCSD(4) and ``order" as it pertains to defining leading order triples corrections to the excited state problem. 

Transcending the baseline level of approximation provided by (e.g. zeroth-order) q-sc-EOM-UCCSD(4) requires us to account for triple excitations effects, requiring $\ket{\textbf{h}_3}$ and 

\begin{equation}\label{eq:HbarUCCSDT4}
\begin{split}
         e^{-\tau_1-\tau_2-\tau_3}\mathcal{H}_{\text{N}}e^{\tau_1+\tau_2+\tau_3}&\approx \mathcal{H}_{\text{UCCSD(4)}} +\underbrace{\bigg( \bar{\mathcal{H}}^{[2]} + \bar{\mathcal{H}}^{[3]}+ \bar{\mathcal{H}}^{[4]} \bigg)}_{\text{$\equiv H_T$}}\\
        &=\mathcal{H} + H_T,
\end{split}
\end{equation} where
\begin{widetext}
\begin{equation}
    \begin{split}
        \bar{\mathcal{H}}^{[2]} &= [H_0,\tau_3]\\
        \bar{\mathcal{H}}^{[3]} &= \frac{1}{2}\bigg([[H_0,\tau_3],\tau_2] +[[H_0,\tau_2],\tau_3]\bigg) + [W_N,\tau_3]\\
        \bar{\mathcal{H}}^{[4]} &= \frac{1}{2}\bigg([[H_0,\tau_3],\tau_3] + [[H_0,\tau_3],\tau_1] +[[H_0,\tau_1],\tau_3]+[[W_N,\tau_2],\tau_3]+[[W_N,\tau_3],\tau_2]\bigg)+\cdots
    \end{split}
\end{equation} 
\end{widetext}
account for commutators involving $\tau_3$. Although $\mathcal{H}_{T}$ contains the second-order contribution
$\overline{\mathcal{H}}^{[2]}=[H_{0},\tau_{3}]$, this term does not contribute to the present second-order excitation-energy functional. In the excitation-only q-sc-EOM representation, $C_{P}=C_{1}+C_{2}$ spans the singles and doubles excitation space, whereas $\overline{\mathcal{H}}^{[2]}$ changes the particle--hole rank by three. Consequently,
\begin{equation}
\langle 0 |
[C_{P}^{\dagger},[\overline{\mathcal{H}}^{[2]},C_{P}]]
|0\rangle = 0 .
\end{equation}
All remaining terms involving $\mathcal{H}_{T}$ either contain the first-order triples operator $C_{3}$ or a higher-order component of $\mathcal{H}_{T}$ and therefore contribute only at third or higher order.

Given Equation \eqref{eq:HbarUCCSDT4}, the expression corresponding to full q-sc-EOM-UCCSDT(4) can be written as 
\begin{equation}\label{eq:fullEOMT}
\begin{split}
        \omega&_{\text{UCCSDT(4)}} \bigg(\braket{0|[C_P^{\dagger},C_P]|0} +\braket{0|[C_3^{\dagger},C_3]|0}\bigg) = \\
        & \braket{0|[C_P^{\dagger},[\mathcal{H},C_P]]|0}+\braket{0|[C_P^{\dagger},[H_T,C_P]]|0} \\
    &+\braket{0|[C_P^{\dagger},[\mathcal{H},C_3]]|0}+\braket{0|[C_P^{\dagger},[H_T,C_3]]|0} \\
    &+\braket{0|[C_3^{\dagger},[\mathcal{H},C_P]]|0}+\braket{0|[C_3^{\dagger},[H_T,C_P]]|0} \\
    &+\braket{0|[C_3^{\dagger},[\mathcal{H},C_3]]|0}+\braket{0|[C_3^{\dagger},[H_T,C_3]]|0} \\
\end{split}
\end{equation} 

Note that Equation \eqref{eq:fullEOMT} implicitly accounts for higher-order triples terms while our immediate focus is on low-order corrections. With this in mind, we only require $\mathcal{H}^{[0]}$ and $\mathcal{H}^{[1]}$ to define the leading, second-order triples correction.  We further assume that $C_1$/$C_2$ (and their respective adjoints) are zeroth-order, and $C_3$ (and its adjoint) would represent a first-order contribution in $W_N$. Isolating all terms of Equation \eqref{eq:fullEOMT} that are second-order in $W_N$ reveals the leading-order functional, $\omega_{[\text{T}]}$, accounting for triple excitation effects:
\begin{equation}\label{eq:secondOfull}
\begin{split}
    \omega&_{[\text{T}]}\left( 1 + \braket{C_3^{\dagger}|C_3}\right) = \omega_{\text{UCCSD}} +\braket{0|[C_3^{\dagger},[H_0,C_3]]|0} \\
    &+ \left(\braket{0|[C_3^{\dagger},[W_N,C_2]]|0}+\braket{0|[C_2^{\dagger},[W_N,C_3]]|0}\right) \\
    &+\left( \braket{0|[C_1^{\dagger},[W_N,C_3]]|0} +\braket{0|[C_1^{\dagger},[[H_0,\tau_2],C_3]]|0}\right)
\end{split}
\end{equation}

\begin{figure}[ht!]
 \centering
 \includegraphics[width=\columnwidth]{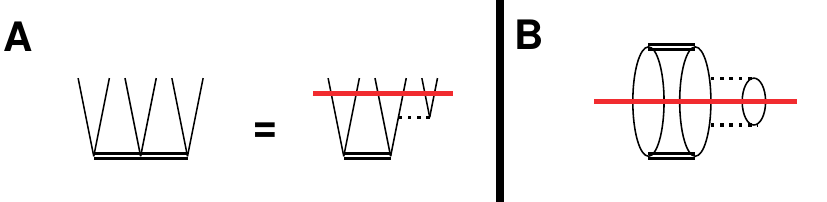}
 \caption{(A) depicts the equations for $C_3$ found in Equation \eqref{eq:TfirstOresidEqns}, while (B) illustrates the [T] triples correction to EOM-UCCSD. Double bar lines denote  $C_2$/$C_2^{\dagger}$ amplitudes. The red horizontal line refers to the $\left(\omega_{UCCSD}+D_3\right)^{-1}$ denominator, where each element of the Fock denominator has been shifted by the q-sc-EOM-UCCSD excitation energy. Dotted lines correspond to elements of $W_N$. }
 \label{fig:trips}
\end{figure}

Note that all higher-order terms of Equation \eqref{eq:HbarUCCSDT4} and those involving $H_T$ from Equation \eqref{eq:fullEOMT} have been discarded in the definition of Equation \eqref{eq:secondOfull}, as they contribute to the energy functional after second-order. Furthermore, we can impose additional simplifications on the functional defined in Equation \eqref{eq:secondOfull} by invoking the UCC(2) ground state residual equation (e.g. MBPT(2) residual equation) such that
\begin{equation}
    \begin{split}
\braket{0|C_1^{\dagger}\underbrace{(T_2^{\dagger}f_N+W)}_{\text{$\approx$0}+\delta(3)}C_3|0}
    \end{split}
\end{equation} For further discussion rationalizing this cancellation, we refer the reader to prior work\cite{bartlett1989some,bartlett1989alternative} -- in particular the first line of Equation S17 of the Supplementary Material to Ref. \cite{windom2024new} which displays this cancellation for the ground state equations. In either case, this cancellation leads to the simplified energy functional 
\begin{equation}\label{eq:secondOsimplified}
\begin{split}
        \omega_{[\text{T}]}&\left( 1 + \braket{C_3^{\dagger}|C_3}\right) = \omega_{\text{UCCSD}} +\braket{0|[C_3^{\dagger},[H_0,C_3]]|0} \\
        &+ \left(\braket{0|[C_3^{\dagger},[W_N,C_2]]|0}+\braket{0|[C_2^{\dagger},[W_N,C_3]]|0}\right)
\end{split}
\end{equation} At this point, we need an associated set of residual equations defining the first-order approximation to $C_3$ by taking the partial derivative, $\frac{\partial}{\partial C_3^{\dagger}}$, of Equation \eqref{eq:secondOsimplified} such that
\begin{equation}\label{eq:TfirstOresidEqns}
    \begin{split}
        &\omega_{\text{[T]}}\braket{\textbf{h}_3|C_3|0} = \braket{\textbf{h}_3|f_NC_3|0}+\braket{\textbf{h}_3|W_NC_2|0} \\
        \Rightarrow&\braket{\textbf{h}_3|C_3|0}\left(\omega_{\text{[T]}}+D_3\right)=\braket{\textbf{h}_3|W_NC_2|0} \\
        \Rightarrow&\braket{\textbf{h}_3|C_3|0}= \left(\omega_{\text{[T]}}+D_3\right)^{-1}\braket{\textbf{h}_3|W_NC_2|0}
    \end{split}
\end{equation} where $D_3\equiv(f_{ii}+f_{jj}+f_{kk}-f_{aa}-f_{bb}-f_{cc})$ is formed with the combination of the 6-index Fock denominator and we adopt the standard convention\cite{watts1996iterative} where $\left(\omega_{\text{[T]}}+D_3\right)^{-1}\approx \left(\omega_{UCCSD}+D_3\right)^{-1}$. Note that the last line of Equation \eqref{eq:TfirstOresidEqns} defining the leading-order triples correction to q-sc-EOM-UCCSD considering $C_3$ is shown diagrammatically in Figure \ref{fig:trips} A.

The solution of this residual equation can be imposed on Equation \eqref{eq:secondOsimplified}, which in turn induces a final simplification on the functional and characterizes the leading, second-order correction to the q-sc-EOM-UCCSD energy:
\begin{widetext}
\begin{equation}
\begin{split}
        \omega_{\text{[T]}} &= \omega_{\text{UCCSD}} +\braket{0|[C_2^{\dagger},[W_N,C_3]]|0} \\
        &= \omega_{\text{UCCSD}}+ \braket{0|(C_2^{\dagger}W_N)_C|\textbf{h}_3}\bigg(\omega_{\text{UCCSD}}+D_3\bigg)^{-1}\braket{\textbf{h}_3|(W_NC_2)_C|0}
\end{split}
\end{equation}
\end{widetext} Henceforth, we refer to this correction as [T], which is shown diagrammatically in  Figure \ref{fig:trips}B and is a direct excited state analog to the corresponding ground state triples correction originally proposed in in Ref. \cite{windom2024new}.

\section{Computational Details}
All computations were performed using the 6-31G basis set using 6 electrons in 6 orbitals active spaces. Molecular integrals and HF reference states were generated with PySCF~\cite{pyscf_2020} and the fermionic Hamiltonians were mapped to qubit operators using the Jordan-Wigner transformation employed by PennyLane's quantum chemistry module~\cite{bergholm2018pennylane}. The energies of the BH and CH$^+$ molecules are calculated at equilibrium bond length ($R_e$), 1.5 $R_e$, and 2.0 $R_e$, and were chosen as representative examples that broadly cover electron correlation regimes.\cite{d2023challenges,kowalski2001active,krylov2000excited} The HF state was used as the initial state in all calculations. Ground state energies and wave functions were obtained using the UCCSD ansatz within the VQE framework. The UCCSD parameters were optimized using the BFGS optimizer on a numerical statevector simulator until the ground-state energy satisfied a convergence threshold of $10^{-8}$ H. The same active space, molecular Hamiltonian, excitation ordering, and optimized UCCSD parameters were then used without further optimization in the q-sc-EOM calculation. The codes generated for all calculations are provided in Ref.~\cite{qcant2025}. The lowest energy eigenstates corresponding to particular spin states of a particular symmetry are studied for the isoelectronic BH and CH$^+$ molecules; specifically, the $^1\Sigma^{+}$, $^3\Sigma^{+}$, $^5\Sigma^{-}$, $^3\Pi$, $^1\Pi$, $^1\Delta$, and $^3\Delta$ electronic states. 

We use two, distinct flavors of UCCSD ans\"{a}tze to determine the set of amplitudes corresponding to the ground state. Specifically, the full UCCSD exponential operator is used, defined as 
\begin{equation}
  \label{eq:uccsd}
  |\Psi_\text{UCCSD}\rangle = e^{\sum_{ia}\theta_{i}^{a}(a^\dagger i- \text{h.c.})+\sum_{ijab}\theta_{ij}^{ab}(a^\dagger b^\dagger ji- \text{h.c.}))} |0\rangle
\end{equation}
as well as a trotterized variant, shown by
\begin{equation}
   |\Psi_\text{tUCCSD}\rangle = \prod_{ia}e^{\theta_i^a(a^\dagger i- \text{h.c.})}\prod_{ijab}e^{\theta_{ij}^{ab}(a^{\dagger}b^{\dagger}ji-\text{h.c.})}|0\rangle
  \label{eq:trotter}
\end{equation}

We note that, while the two ans\"{a}tze in Equations \eqref{eq:uccsd} and \eqref{eq:trotter} share similarities, recent literature suggests that the trotterized version is its own unique ansatz construct.\cite{evangelista2019exact, Freericks2022} In both cases, we leverage the VQE algorithm to variationally optimize the set of cluster amplitudes according to the criteria: 
\begin{equation}
 \tau_1^*,\tau_2^* = \argminA_{ \tau_1,\tau_2}\braket{\Psi(\tau_1,\tau_2)| H | \Psi( \tau_1,\tau_2)},
\end{equation}
which are subsequently decomposed into the corresponding $T$ and $T^\dagger$ and inserted into the equations derived in Sec. \ref{sec:theory}.

\section{Results}

As the magnitude of double excitation character is a useful proxy to ascertain the necessity of higher-order operators, the following subsections provide illustrations that  decompose a net excitation into pure single (i.e., \% $C_1$) and double (i.e., \% $C_2$) components for the aforementioned electronic states. This analysis relies on the normalization condition  
\begin{equation}\label{eq:percentR}
   \mathcal{C}^{\dagger}\mathcal{C} = 1 = \sum_{n\in\{i,a\}} |C_{1,n}|^2 +\sum_{n\in\{i,j,a,b\}} |C_{2,n}|^2
\end{equation} established in q-sc-EOM-t/UCCSD, where each sum goes over each element of the $C_1$ and $C_2$ tensors. Consequently, the percent contribution of $\mathcal{C}$ arising from pure single excitations is given by the first component of Equation \eqref{eq:percentR} whereas the percent contribution  arising from pure doubles excitations is given by the second component of  Equation \eqref{eq:percentR}. The expectation is that states dominated by singles $C_1$ effects will be well-described within baseline q-sc-EOM-t/UCCSD while states involving higher contributions from $C_2$---indicative of biexcitation character---will demand description of higher-order excitation effects as provided by the [T] correction. Note that in what follows, we illustrate the decomposition of Equation \eqref{eq:percentR} with respect to the full q-sc-EOM-UCCSD as we found that the trotterized variant was not significantly different.  

\subsection{BH}

\begin{figure*}[t!]
\centering
\includegraphics[width=0.75\textwidth]{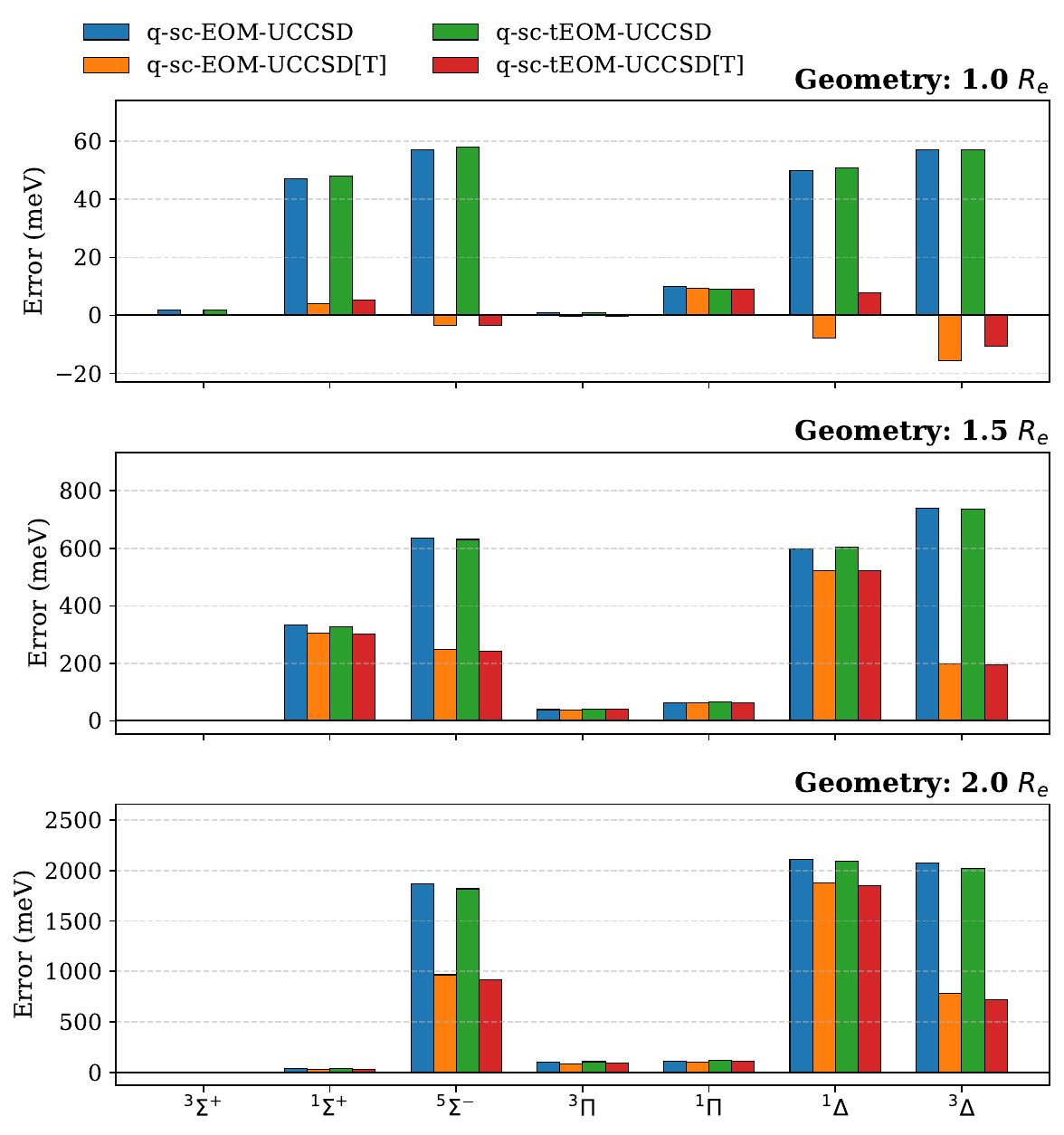}    

\caption{Depiction of the error with respect to the CASCI solution for the lowest $^3\Sigma^{+}$, $^1\Sigma^{+}$, $^5\Sigma^{+}$, $^3\Pi$, $^1\Pi$, $^1\Delta$ and $^3\Delta$ electronically excited states at three different geometries for the BH molecule.}
\label{fig:BHerror}
\end{figure*}

Excited state energy predictions for the BH molecule for all three geometries are recorded in Table \ref{tab:BHEEtab}, and all errors are illustrated in Figure \ref{fig:BHerror}.

\begin{table}[htbp]
\centering
\caption{Excitation energies (eV) for various electronic states of the BH molecule across three geometries, calculated using CASCI and variants of q-sc-EOM-UCCSD.}
\label{tab:BHEEtab}
\small
\begin{tabular}{llccccc}
\toprule
\textbf{Geom.} & \textbf{State} & \textbf{CASCI} & \textbf{UCCSD} & \textbf{[T]} & \textbf{tUCCSD} & \textbf{t[T]} \\\midrule
\multirow{7}{*}{1.0 $R_e$} 
 &  $^3\Sigma^{+}$ & 8.402	&8.404	&8.402	&8.404	&8.402 \\
 &  $^1\Sigma^{+}$ & 8.111	&8.158	&8.115	&8.159	&8.116\\
 &  $^5\Sigma^{-}$ & 8.161	&8.218	&8.157	&8.219	&8.158 \\
 &  $^3\Pi$        & 1.300	&1.301	&1.300	&1.301	&1.300 \\
 &  $^1\Pi$        & 3.376	&3.386	&3.385	&3.385	&3.385 \\
 &  $^1\Delta$     & 6.957	&7.007	&6.949	&7.008	&6.965 \\
 &  $^3\Delta$     & 11.903	&11.960	&11.888	&11.960	&11.892\\
\midrule
\multirow{7}{*}{1.5 $R_e$} 
 &  $^3\Sigma^{+}$ & 3.691	&3.694	&3.691	&3.693	&3.690 \\
 &  $^1\Sigma^{+}$ & 8.109	&8.441	&8.413	&8.437	&8.410 \\
 &  $^5\Sigma^{-}$ &5.815	&6.450	&6.062	&6.446	&6.058\\
 &  $^3\Pi$        & 1.660	&1.699	&1.698	&1.701	&1.699 \\
 &  $^1\Pi$        & 2.518	&2.581	&2.579	&2.583	&2.581 \\
 &  $^1\Delta$     & 7.406	&8.004	&7.929	&8.009	&7.930 \\
 &  $^3\Delta$     & 9.509	&10.250	&9.707	&10.246	&9.702 \\
\midrule
\multirow{7}{*}{2.0 $R_e$} 
 &  $^3\Sigma^{+}$ & 1.265	&1.263&	1.261&	1.263&	1.262 \\
 &  $^1\Sigma^{+}$ & 6.136	&6.173&	6.167&	6.173&	6.167 \\
 &  $^5\Sigma^{-}$ & 4.297	&6.169&	5.264&	6.117&	5.214 \\
 &  $^3\Pi$        & 0.888	&0.986&	0.973&	0.993&	0.980 \\
 &  $^1\Pi$        & 1.027	&1.139&	1.126&	1.146&	1.133 \\
 &  $^1\Delta$     & 7.371	&9.482&	9.249&	9.469&	9.227 \\
 &  $^3\Delta$     & 7.928	&10.007&8.708&	9.952&	8.645 \\
\bottomrule
\end{tabular}
\end{table}

\begin{figure*}[ht!]
\centering
\includegraphics[width=0.75\textwidth]{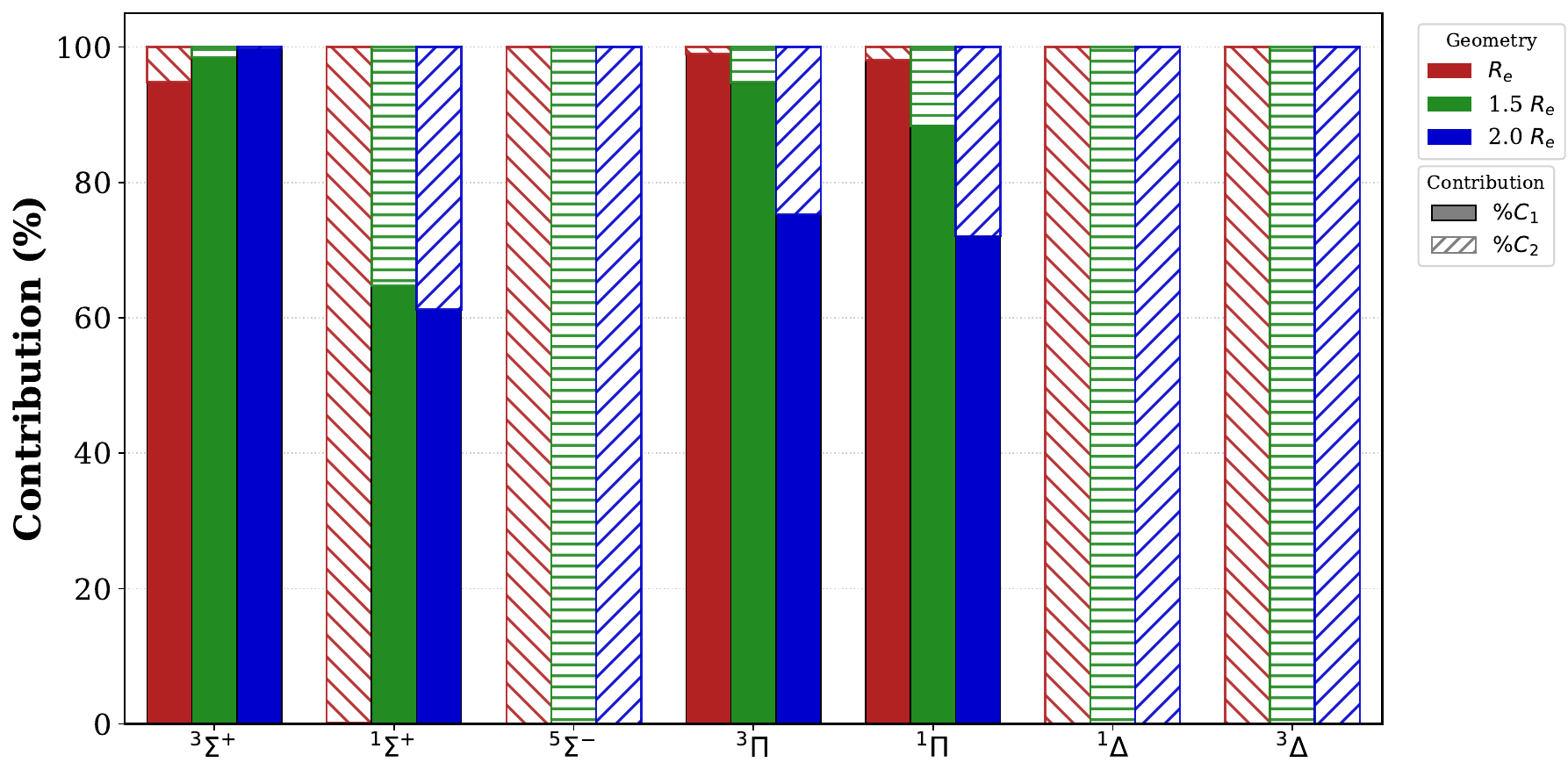}    

\caption{Decomposition of the q-sc-EOM-UCCSD eigenvectors in terms of their contribution from $C_1$ and $C_2$ for the BH molecule in three different geometries.}
\label{fig:BHpercentR}
\end{figure*}

Based on the contribution from singles and doubles to the q-sc-EOM-t/UCCSD excitation vectors (Figure \ref{fig:BHpercentR}),  the lowest energy $^3\Sigma^{+}$ state is dominated by singles excitations at all geometries. Analysis of the dominant determinants in the CASCI solution indicate transitions of the form $3\sigma \rightarrow 4\sigma$ are most prominent. Because of this,  q-sc-EOM-t/UCCSD adequately represents this excited state incurring errors $\approx$1-3 meV for all geometries. The [T] correction offers modest improvements, although results slightly worsen when using the both trotterized and full [T] at 2.0 $R_e$. We conjecture this is influenced by the fact that the baseline q-sc-t/EOM-UCCSD result at 2.0 $R_e$ slightly underpredicts the CASCI solution, which in turn negatively biases the corresponding [T] correction. 


On the other hand, the $^1\Sigma^{+}$ excited state is dominated by double excitations of the form 3$\sigma^2 \rightarrow 1\pi^2$ in the CASCI solution at $R_e$, progressing to mixed doubly and singly excited configurations like $3\sigma^2 \rightarrow 2\pi^2$ and $3\sigma \rightarrow 4\sigma$, respectively, in stretched regimes. This behavior is well-emulated by  Figure \ref{fig:BHpercentR}. However, q-sc-EOM-t/UCCSD exhibits an error margin that is quite extreme for this state. For example, both full and trotterized variants of q-sc-EOM-UCCSD exhibit errors on the order of 48 meV and 37 meV at $R_e$ and 2.0 $R_e$, respectively. The [T] correction reduces this error to $\approx$5 meV at $R_e$, whereas the magnitude of improvement is more modest at 2.0 $R_e$---representing a reduction by about $\approx$6 meV for both ansatz. On the other hand, the error of q-sc-EOM-t/UCCSD at 1.5 $R_e$ is quite large at about 330 meV; in this instance, the [T] improves agreement with the CASCI solution by about 27 meV for full/trotterized ansatze.  

As opposed to the other multiplet states, the $^5\Sigma^{-}$ state is dominated by $C_2$ effects for all geometries, with configurations like $2\sigma3\sigma\rightarrow 1\pi_x1\pi_y$  participating significantly in the CASCI solution. The benefits offered by the [T] correction for this electronic state are  readily apparent. In this case, the [T] correction significantly mitigates the error by roughly 61 meV and ~388 meV at $R_e$ and 1.5 $R_e$, respectively, as compared to baseline q-sc-EOM-t/UCCSD. The magnitude of improvement provided by [T] is even more extreme at 2.0 $R_e$, reducing the error of baseline q-sc-EOM-t/UCCSD by over 900 meV. Although it should be recognized that the prediction of both baseline q-sc-EOM-t/UCCSD and the [T] correction  nevertheless fall short of being ``chemically accurate" at 1.5 $R_e$ and 2.0 $R_e$, suggesting higher-order corrections may be warranted for this electronic state.

Both $^3\Pi$ and $^1\Pi$ excited states are dominated by single excitations at every geometry, as evident from Figure \ref{fig:BHpercentR}. The magnitude of this effect is most evident at $R_e$, which gradually diminishes at stretched geometries where $C_2$ effects begin to participate to some extent. As a consequence, baseline q-sc-EOM-t/UCCSD incurs an error close to 1 and 10 meV for these states at $R_e$, respectively; in both cases for both ansatz, the [T] correction reduces the error by roughly 0.5-1 meV. This trend continues for 1.5 $R_e$, where both operator-versions of [T] correction reduces the error of baseline q-sc-EOM-UCCSD by roughly 1-2 meV for the $^3\Pi$ and $^1\Pi$ excited states. This magnitude of improvement offered by [T] increases slightly at 2.0 $R_e$, where  agreement with the CASCI solution is improved by $\approx$ 12-14 meV for both ansatze and both $^3\Pi$ and $^1\Pi$ excited states. 

Figure \ref{fig:BHpercentR} clearly demonstrates that the $^1\Delta$ and $^3\Delta$ excited states are dominated by biexcitation effects, alluding to the importance of accounting for triple excitation effects. As revealed by Table \ref{tab:BHEEtab}, the magnitude of improvement offered by q-sc-EOM-UCCSD[T] is slightly larger than the trotterized [T] correction for both $^1\Delta$ and $^3\Delta$ excited states; regardless, full (trotterized) [T] represents a 58/72 ( 43/68 ) meV improvement over the baseline q-sc-EOM-(t)UCCSD  result for the $^1\Delta$/$^3\Delta$ states, respectively. For both ansatze at 1.5$R_e$, the magnitude of the [T] serves as a $>$70/540 meV for the $^1\Delta$/$^3\Delta$ states as compared to the baseline. The extent of this improvement is maximal at 2.0$R_e$, where [T] improves agreement with the CASCI solution by $>$230/1290 meV for the $^1\Delta$/$^3\Delta$ states as compared to the baseline.Despite these impressive levels of improvement, the [T] correction is nevertheless unable to adequately resolve the $^1\Delta$ and $^3\Delta$ excited states to acceptable precision despite representing a significant improvement over baseline q-sc-EOM-t/UCCSD results. This, in turn, is indicative of  the importance of higher-order excitation effects for these electronic states in particular. 


In general, we note that prediction errors increase at 1.5 $R_e$ and become most acute overall at 2.0 $R_e$; though, this is not surprising as electronic correlations are expected to become stronger in these stretched regimes. Indeed, the errors of both $^1\Delta$ and $^3\Delta$ become extreme at these configurations. However, inspection of the dominant determinants participating in the CASCI expansion reveal increasing participation of triply excited determinant. For example,  contributions arising from $2\sigma3\sigma^2 \rightarrow 1\pi_x1\pi_y4\sigma$ transitions begin to participate in the $^3\Delta$ electronic state at 2.0 $R_e$. Of course, electronic states mediated by triple excitation effect require a theory that adequately captures quadruple excitation effects, thereby explaining the severity of error for states of this type. On the other hand, electronic states like $^3\Sigma^{+}$ and $^1\Sigma^{+}$ are exceptions to this analysis, as single excitation effects become more pronounced at these stretched geometries which can be inferred by Figure \ref{fig:BHpercentR}. Analysis of the dominant CASCI determinant weights corroborate this; the dominant coefficients mediate transitions like $3\sigma \rightarrow 4\sigma$ in the $^3\Sigma^{+}$ state, for example.

\subsection{CH$^+$}
\begin{figure*}[ht!]
\centering
\includegraphics[width=0.75\textwidth]{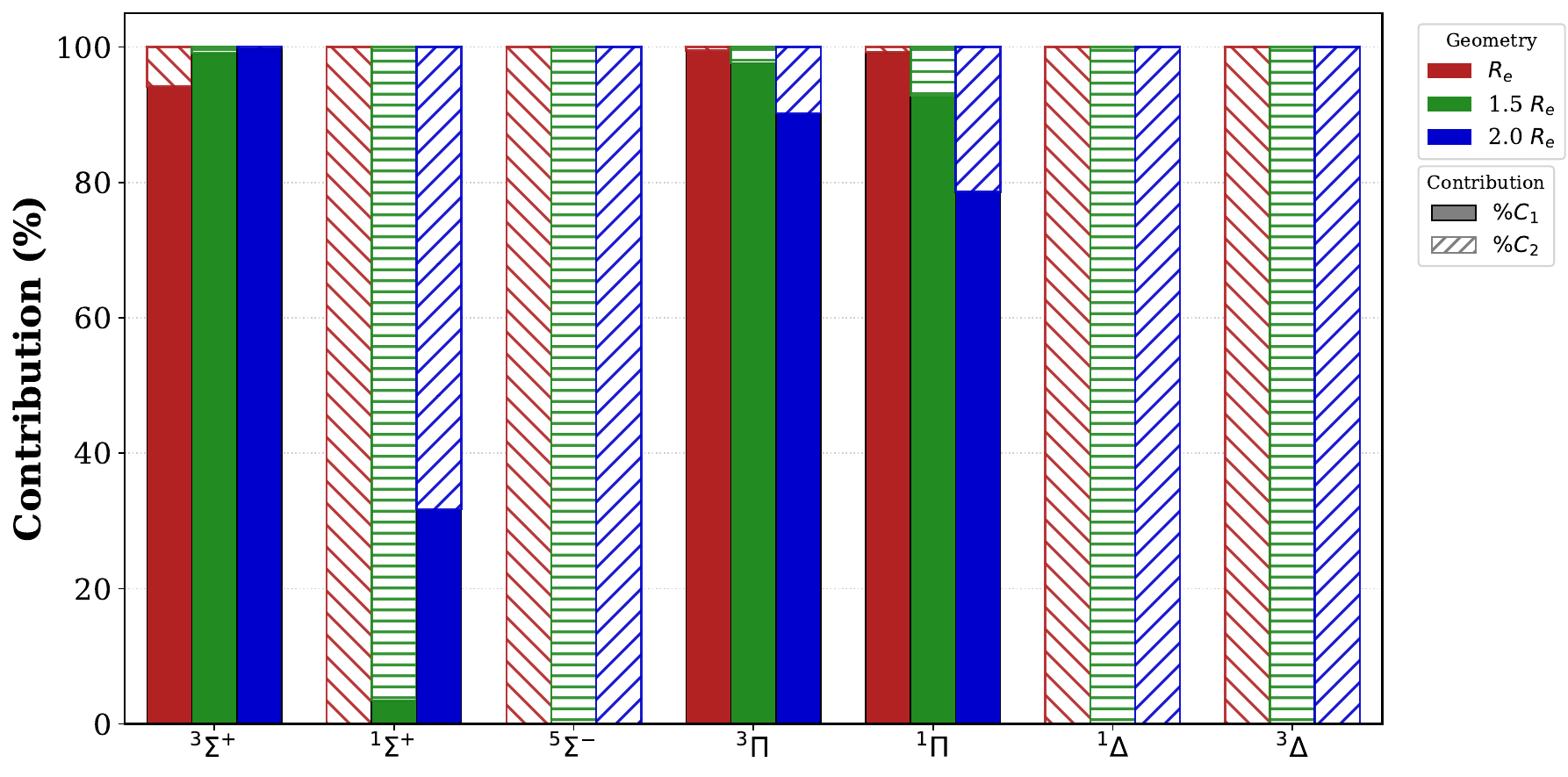}    

\caption{Decomposition of the q-sc-EOM-UCCSD eigenvectors in terms of their contribution from $C_1$ and $C_2$ for the CH$^+$ molecule in three different geometries.}
\label{fig:CHpercentR}
\end{figure*}

Excited state energy predictions for the CH$^+$ molecule for all three geometries are recorded in Table \ref{tab:CH$^+$EEtab}. The choice to include the lowest energy $^1\Sigma^{+}$, $^1\Pi$, and $^1\Delta$ electronic states in the current analysis was informed by prior work.\cite{kowalski2001active,krylov2000excited} Based on Figure \ref{fig:CHpercentR}, only the $^3\Sigma^{+}$, $^1\Pi$, and $^3\Pi$ electronic states are dominated by singles excitations (e.g. $>$75\% $C_1$ excitation character). Consequently, we expect the remaining excited states to be characterized by significant two electron-mediated excitation effects, which can be broadly recognized upon inspection of the results recorded in Table \ref{tab:CH$^+$EEtab}. Results for CH$^+$ are arguably more distributed overall than those for BH. We also note that the molecular orbital symmetry abruptly changes at 2.0 $R_e$ for CH$^+$, where the lowest to highest energy virtual orbital assignment transitions from B$_1$, B$_2$, A$_1$ at 1.0 $R_e$ and 1.5 $R_e$ to A$_1$, B$_1$, B$_2$.

\begin{table}[htbp]
\centering
\caption{Excitation energies (eV) for various electronic states of the CH$^+$ molecule across three geometries, calculated using CASCI and variants of q-sc-EOM-UCCSD.}
\label{tab:CH$^+$EEtab}
\small
\begin{tabular}{llccccc}
\toprule
\textbf{Geom.} & \textbf{State} & \textbf{CASCI} & \textbf{UCCSD} & \textbf{[T]} & \textbf{tUCCSD} & \textbf{t[T]} \\\midrule
\multirow{7}{*}{1.0 $R_e$} 
 &  $^3\Sigma^{+}$ & 11.764	&11.766	&11.763&	11.766&	11.763\\
 &  $^1\Sigma^{+}$ & 8.946	&8.956	&8.950	&8.956	&8.950 \\
 &  $^5\Sigma^{-}$ & 9.070	&9.077	&9.014	&9.076	&9.014 \\
 &  $^3\Pi$        & 1.289	&1.293	&1.293	&1.293	&1.292 \\
 &  $^1\Pi$        & 3.568	&3.575	&3.575	&3.575	&3.575 \\
 &  $^1\Delta$     & 7.691	&7.699&	7.682	&7.699	&7.693 \\
 &  $^3\Delta$     & 14.144	&14.144&14.049	&14.145	&14.055 \\
\midrule
\multirow{7}{*}{1.5 $R_e$} 
 &  $^3\Sigma^{+}$ & 4.140	&4.140&	4.138	&4.140&	4.138 \\
 &  $^1\Sigma^{+}$ & 7.227	&7.347&	7.315&	7.350&	7.318 \\
 &  $^5\Sigma^{-}$ & 6.574	&6.579&	6.357&	6.582&	6.359 \\
 &  $^3\Pi$        & 1.116	&1.157&	1.155&	1.157&	1.155 \\
 &  $^1\Pi$        & 2.197	&2.258	&2.254&	2.258&	2.254 \\
 &  $^1\Delta$     & 6.401	&6.540	&6.488&	6.546&	6.493\\
 &  $^3\Delta$     & 11.621	&11.669	&11.325&	11.674&	11.331 \\
\midrule
\multirow{7}{*}{2.0 $R_e$} 
 &  $^3\Sigma^{+}$ & 1.185&	1.181&	1.181&	1.182	&1.182 \\
 &  $^1\Sigma^{+}$ & 5.781	&5.888&	5.857&	5.887&	5.857 \\
 &  $^5\Sigma^{-}$ & 5.970&	5.969&	5.710&	5.969&	5.711 \\
 &  $^3\Pi$        & 0.818&	0.949&	0.932&	0.937&	0.918 \\
 &  $^1\Pi$        & 1.188&	1.344&	1.321&	1.330&	1.306 \\
 &  $^1\Delta$     & 5.478&	5.756&	5.678&	5.755&	5.678 \\
 &  $^3\Delta$     & 10.942&	10.970&	10.548&	10.982&	10.561 \\
\bottomrule
\end{tabular}
\end{table}

\begin{figure*}[ht!]
\centering
\includegraphics[width=0.75\textwidth]{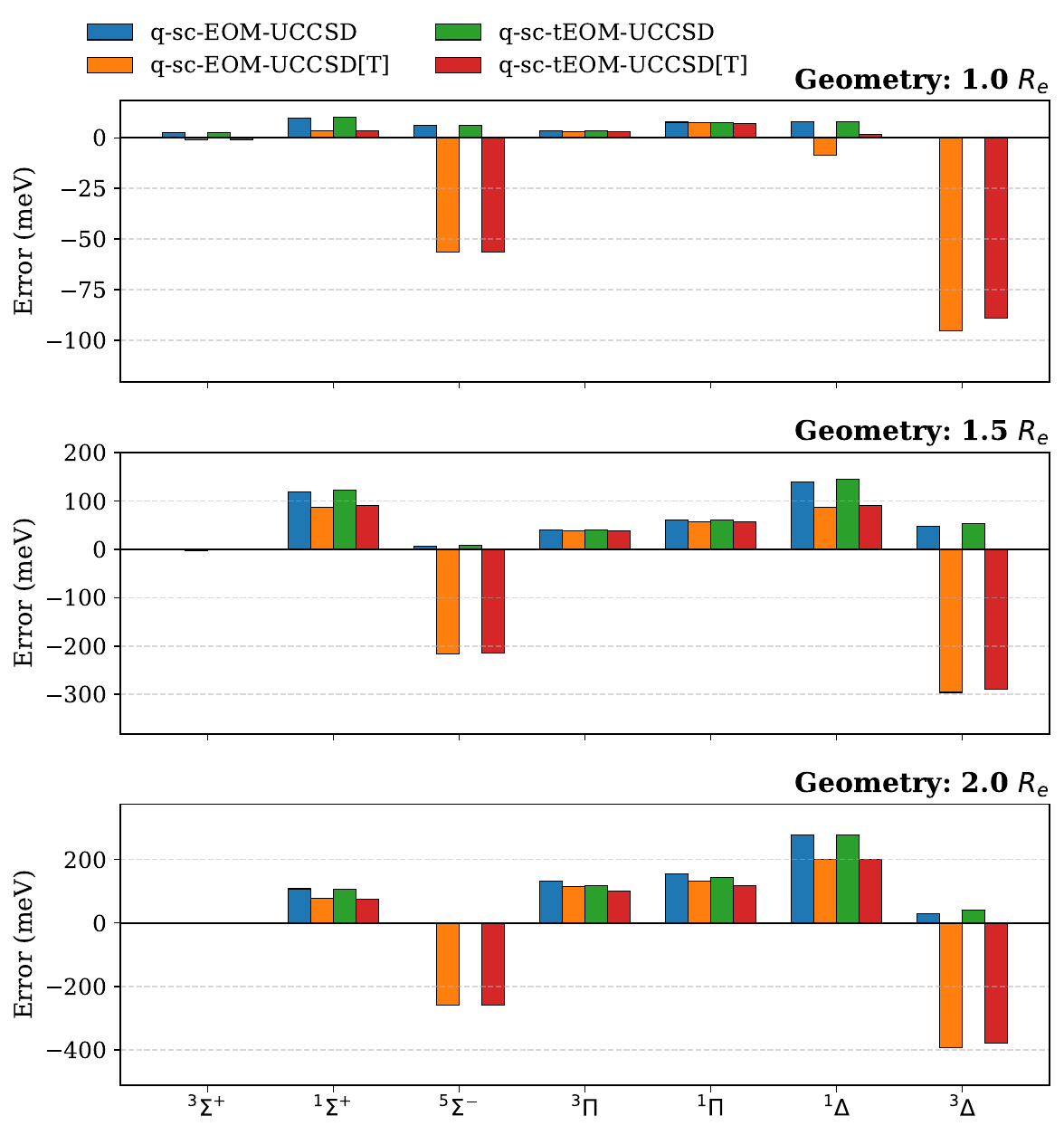}    

\caption{Depiction of the error with respect to the CASCI solution for the lowest $^3\Sigma^{+}$, $^1\Sigma^{+}$, $^5\Sigma^{+}$, $^3\Pi$, $^1\Pi$, $^1\Delta$ and $^3\Delta$ electronically excited states at three different geometries for the CH$^+$ molecule.}
\label{fig:CH$^+$error}
\end{figure*}

The errors in excitation energies with respect to CASCI are plotted in Figure \ref{fig:CH$^+$error}. Specifically, the magnitude of  the [T] error in the $^5\Sigma^{-}$ and $^3\Delta$ states are quite large across all geometries, tending to significantly underestimate both the CASCI and q-sc-EOM-t/UCCSD result. Figure \ref{fig:CHpercentR} highlights both electronic states being 100\% C$_2$ dominated, which is also the case for BH. Inspection of the dominant CASCI determinant reveals biexcitations facilitating $2\sigma3\sigma\rightarrow 1\pi^2$ and $2\sigma3\sigma\rightarrow 1\pi_x1\pi_y$ transitions for the $^3\Delta$ state. Similarly, various $2\sigma3\sigma\rightarrow 1\pi_x1\pi_y$ transitions are present in the CASCI solution for the $^5\Sigma^{-}$ state. Curiously, the baseline q-sc-EOM prediction for these states is better for CH$^+$ than for BH, yet the corresponding [T] correction leads to significantly poorer  results for CH$^+$. In fact, [T] significantly underpredicts the CASCI solution for these states across all CH$^+$ geometries. Further inspection of the maximal $C_2$ and $T_2$ amplitudes for the $^5\Sigma^{-}$ and $^3\Delta$ states of CH$^+$ indicate significant multireference effects (e.g. prescence of ground \textbf{and} excited state amplitudes exceeding a value of  0.4). This suggests that the assumptions of the underlying perturbation theory may no longer be applicable when describing these electronic states.  Regardless, in all cases where the baseline q-sc-EOM-t/UCCSD underpredicts the CASCI result, the corresponding [T] correction --- at best --- offers no improvement, and at worst exacerbates the underlying error.

On the other hand, [T] results for the remaining biexcited-dominated states---$^1\Sigma^{+}$ and $^1\Delta$---represent an overall improvement over baseline q-sc-EOM. Starting with $^1\Sigma^{+}$, examination of the dominant CASCI coefficients reveal $3\sigma^2 \rightarrow 1\pi^2$ excitations dominating at $R_e$ and 1.5 $R_e$, with HOMO to LUMO (e.g. A$_1$ to A$_1$) single excitations making a finite contribution at 2.0 $R_e$, which might be assume from analysis of Figure \ref{fig:CHpercentR}. At $R_e$, the [T] correction cuts the error of baseline q-sc-EOM-t/UCCSD by more than half -- $\approx$ 6 meV. At 1.5 and 2.0 $R_e$, the [T] correction reduces the error of baseline q-sc-EOM-t/UCCSD by $\approx$ 30 meV. 

For the $^1\Delta$ state, analysis of the CASCI solution reveals that $3\sigma^2 \rightarrow 1\pi^2$ transition dominates at $R_e$ with limited weight supporting $3\sigma^2\rightarrow 1\pi_x1\pi_y$ transitions. The importance of the $3\sigma^2\rightarrow 1\pi_x1\pi_y$ transition increases in the CASCI solution at 1.5 $R_e$, eventually representing a dominant contribution at 2.0 $R_e$. The trotterized [T] improves the error by 6 meV as compared to the baseline at $R_e$; however the full [T] variant ends up underpredicting the CASCI solution, and has an absolute error magnitude that is roughly the same as ordinary q-sc-EOM-UCCSD. However, the [T] correction improves upon baseline q-sc-EOM-t/UCCSD by $>$ 50 meV and 77 meV at 1.5 $R_e$ and 2.0 $R_e$, respectively. Although the magnitude of improvement offered by [T] is significant for these examples, it is worth pointing out that the errors experienced by all methods remains quite large.

The remaining $^3\Sigma^{+}$, $^3\Pi$, and $^1\Pi$ electronic states are all singles-dominated excited states. As shown in Figure \ref{fig:CHpercentR},  the preeminence of single excitation effects arising in the $^3\Sigma^{+}$ state increases once the CH$^+$ bond is stretched; the opposite is true for the $^3\Pi$ and $^1\Pi$ states. Regarding the $^3\Sigma^{+}$ state, both operator variants of [T] corrrection do not meaningfully modify the baseline q-sc-EOM-t/UCCSD result in absolute terms. Improvements offered by [T] tend to be more impactful for the $^3\Pi$ and $^1\Pi$ states at stretched geometries. Both operator variants of [T] offer modest improvements of $\approx$ 2/3 meV over the baseline q-sc-EOM  prediction for the $^3\Pi$/$^1\Pi$ states at 1.5 $R_e$. The significance of this trend increases slightly at 2.0$R_e$, where [T] improves over baseline q-sc-EOM-UCC by $\approx$ 18/23 meV for the $^3\Pi$/$^1\Pi$ states, respectively.

\section{Conclusion}
We propose a leading-order correction for neglected triple excitations in q-sc-EOM-UCCSD theory, which we refer to as [T]. A hybrid compute paradigm is leveraged, wherein the [T] correction is computed using a $\mathcal{O}(N^7)$ algorithm that only requires classical resources. We emphasize the [T] correction requires no quantum resources beyond baseline q-sc-t/EOM-UCCSD. The utility of this correction is interrogated in a benchmark study of doubly (and singly) excited states using both the trotterized and full operator variants of q-sc-EOM-UCCSD. Our findings suggest that the improvements offered by [T] can be quite large, particularly for excited states exhibiting significant double excitation character. Across the present benchmark, the [T] correction generally reduces excitation-energy errors when the underlying q-sc-EOM-UCCSD reference remains perturbatively well behaved. Large improvements are observed for several states with substantial double-excitation character. Two notable exceptions to this trend were found in CH$^+$, where [T] was found to be highly sensitive to multireference effects leading to overestimation of the triples contribution which in turn increases the error.


Our analysis demonstrates that, overall, the merits of the [T] correction outweigh its deficiencies. The [T] correction offers an attractive and cost-efficient procedure to other alternatives that have previously been suggested to capture triple excitation effects, like those based on higher-order RDMs which might require excessive quantum resources.  Triples corrections incorporating measured RDMs are not guaranteed to be size-intensive since they contain both cumulant and non-cumulant terms---meaning that  PT corrections designed in this fashion will potentially  sacrifice an exact condition satisfied by an exact theory. Furthermore, RDM-based triples corrections demand enormous overhead, with recent work\cite{zheng2024quantum} suggesting $\mathcal{O}(N_o^5N_v^5)\sim \mathcal{O}(N^{10})$ measurements would be required for higher-order RDM construction; this is in stark contrast to the purely classical $\mathcal{O}(N^{7})$ algorithm required to compute [T] without any addition of any quantum resources. On the other hand, it is possible that RDM-based triples corrections might be more  robust in stretched regimes where [T] is more susceptible to catastrophic failure. Future work to quantify this is certainly warranted.

\section{Acknowledgments}
 Z.W.W. and D.C. acknowledge support by the “Embedding Quantum Computing into Many-body Frameworks for Strongly Correlated Molecular and Materials Systems” project, which is funded by the U.S. Department of Energy (DOE), Office of Science, Office of Basic Energy Sciences, the Division of Chemical Sciences, Geosciences, and Biosciences. L.B. acknowledges support from the Laboratory Directed Research and Development Program of Oak Ridge National Laboratory, managed by UT-Battelle, LLC, for the US Department
of Energy. R.J.B acknowledges support from the Air Force Office of Scientific Research under AFOSR Award No. FA9550-23-1-0118. P.F.K. and A.A. are funded by NSF grant number 2427046. This research used resources of the Compute and Data Environment for Science (CADES) at the Oak Ridge National Laboratory, which is supported by the Office of Science of the U.S. Department of Energy under Contract No. DE-AC05-00OR22725.

\section{Data Availability Statement}
The data that supports the findings of this study are available within the article and generated using previously disclosed software.\cite{qcant2025}

\bibliography{biblio}

@article{bergholm2018pennylane,
  title={Pennylane: Automatic differentiation of hybrid quantum-classical computations},
  author={Bergholm, Ville and Izaac, Josh and Schuld, Maria and Gogolin, Christian and Ahmed, Shahnawaz and Ajith, Vishnu and Alam, M Sohaib and Alonso-Linaje, Guillermo and AkashNarayanan, B and Asadi, Ali and others},
  journal={arXiv preprint arXiv:1811.04968},
  year={2018}
}

@misc{qcant2025,
  author       = {Sundar, Srivathsan P. and Kwao, Prince and Asthana, Ayush},
  title        = {{QCANT}: Utilities for near-term applications of quantum computing in chemistry and materials science},
  year         = {2025},
  version      = {1.0.0},
  howpublished = {\url{https://github.com/asthanalab/QCANT}}
}

@article{Pyscf_2020,
author = {Sun,Qiming  and Zhang,Xing  and Banerjee,Samragni  and Bao,Peng  and Barbry,Marc  and Blunt,Nick S.  and Bogdanov,Nikolay A.  and Booth,George H.  and Chen,Jia  and Cui,Zhi-Hao  and Eriksen,Janus J.  and Gao,Yang  and Guo,Sheng  and Hermann,Jan  and Hermes,Matthew R.  and Koh,Kevin  and Koval,Peter  and Lehtola,Susi  and Li,Zhendong  and Liu,Junzi  and Mardirossian,Narbe  and McClain,James D.  and Motta,Mario  and Mussard,Bastien  and Pham,Hung Q.  and Pulkin,Artem  and Purwanto,Wirawan  and Robinson,Paul J.  and Ronca,Enrico  and Sayfutyarova,Elvira R.  and Scheurer,Maximilian  and Schurkus,Henry F.  and Smith,James E. T.  and Sun,Chong  and Sun,Shi-Ning  and Upadhyay,Shiv  and Wagner,Lucas K.  and Wang,Xiao  and White,Alec  and Whitfield,James Daniel  and Williamson,Mark J.  and Wouters,Sebastian  and Yang,Jun  and Yu,Jason M.  and Zhu,Tianyu  and Berkelbach,Timothy C.  and Sharma,Sandeep  and Sokolov,Alexander Yu.  and Chan,Garnet Kin-Lic },
title = {Recent developments in the {PySCF} program package},
journal = {The Journal of Chemical Physics},
volume = {153},
number = {2},
pages = {024109},
year = {2020},
}

@article{prasad1985some,
  title={Some aspects of self-consistent propagator theories},
  author={Prasad, M Durga and Pal, Sourav and Mukherjee, Debashis},
  journal={Physical Review A},
  volume={31},
  number={3},
  pages={1287},
  year={1985},
  publisher={APS}
}

@article{szekeres2001killer,
  title={On the “killer condition” in the equation-of-motion method: ionization potentials from multi-reference wave functions},
  author={Szekeres, Zsolt and Szabados, {\'A}gnes and K{\'a}llay, Mih{\'a}ly and Surj{\'a}n, P{\'e}ter R},
  journal={Physical Chemistry Chemical Physics},
  volume={3},
  number={5},
  pages={696--701},
  year={2001},
  publisher={Royal Society of Chemistry}
}

@article{mertins1996algebraic,
  title={Algebraic propagator approaches and intermediate-state representations. {II}. {The equation-of-motion methods for $N$, $N\pm$1, and $N\pm$2 electrons}},
  author={Mertins, F and Schirmer, J and Tarantelli, A},
  journal={Physical Review A},
  volume={53},
  number={4},
  pages={2153},
  year={1996},
  publisher={APS}
}

@article{weiner1980calculation,
  title={Calculation of optimal generalized antisymmetrized geminal-power (projected—bardeen-cooper-schrieffer) functions and their associated excitation spectrum},
  author={Weiner, Brian and Goscinski, Osvaldo},
  journal={Physical Review A},
  volume={22},
  number={6},
  pages={2374},
  year={1980},
  publisher={APS}
}

@article{hodecker2020unitary,
  title={Unitary coupled cluster ground-and excited-state molecular properties},
  author={Hodecker, Manuel and Dreuw, Andreas},
  journal={The Journal of Chemical Physics},
  volume={153},
  number={8},
  year={2020},
  publisher={AIP Publishing}
}

@article{datta1993consistent,
  title={Consistent propagator theory based on the extended coupled-cluster parametrization of the ground state},
  author={Datta, Barnali and Mukhopadhyay, Debasis and Mukherjee, Debashis},
  journal={Physical Review A},
  volume={47},
  number={5},
  pages={3632},
  year={1993},
  publisher={APS}
}

@article{kowalski2001active,
  title={The active-space equation-of-motion coupled-cluster methods for excited electronic states: {Full EOMCCSDt}},
  author={Kowalski, Karol and Piecuch, Piotr},
  journal={The Journal of Chemical Physics},
  volume={115},
  number={2},
  pages={643--651},
  year={2001},
  publisher={American Institute of Physics}
}

@article{krylov2000excited,
  title={Excited states theory for optimized orbitals and valence optimized orbitals coupled-cluster doubles models},
  author={Krylov, Anna I and Sherrill, C David and Head-Gordon, Martin},
  journal={The Journal of Chemical Physics},
  volume={113},
  number={16},
  pages={6509--6527},
  year={2000},
  publisher={American Institute of Physics}
}

@article{asthana2023quantum,
  title={Quantum self-consistent equation-of-motion method for computing molecular excitation energies, ionization potentials, and electron affinities on a quantum computer},
  author={Asthana, Ayush and Kumar, Ashutosh and Abraham, Vibin and Grimsley, Harper and Zhang, Yu and Cincio, Lukasz and Tretiak, Sergei and Dub, Pavel A and Economou, Sophia E and Barnes, Edwin and Mayhall, Nicholas J},
  journal={Chemical Science},
  volume={14},
  number={9},
  pages={2405--2418},
  year={2023},
  publisher={Royal Society of Chemistry}
}

@article{ollitrault2020quantum,
  title={Quantum equation of motion for computing molecular excitation energies on a noisy quantum processor},
  author={Ollitrault, Pauline J and Kandala, Abhinav and Chen, Chun-Fu and Barkoutsos, Panagiotis Kl and Mezzacapo, Antonio and Pistoia, Marco and Sheldon, Sarah and Woerner, Stefan and Gambetta, Jay M and Tavernelli, Ivano},
  journal={Physical Review Research},
  volume={2},
  number={4},
  pages={043140},
  year={2020},
  publisher={APS}
}

@article{liu2022quadratic,
  title={Quadratic unitary coupled-cluster singles and doubles scheme: Efficient implementation, benchmark study, and formulation of an extended version},
  author={Liu, Junzi and Matthews, Devin A and Cheng, Lan},
  journal={Journal of Chemical Theory and Computation},
  volume={18},
  number={4},
  pages={2281--2291},
  year={2022},
  publisher={ACS Publications}
}

@article{windom2025toward,
  title = {Toward the “platinum standard” of quantum chemistry on quantum computers: Perturbative quadruple corrections in unitary coupled cluster theory},
  volume = {164},
  ISSN = {1089-7690},
  number = {1},
  journal = {The Journal of Chemical Physics},
  publisher = {AIP Publishing},
  author = {Windom,  Zachary W. and Bertels,  Luke and Claudino,  Daniel and Bartlett,  Rodney J.},
  year = {2026},
  month = jan,
  pages=014102
}

@article{greenediniz2021generalized,
  author  = {Greene-Diniz, Gabriel and Mu{\~n}oz Ramo, David},
  title   = {Generalized Unitary Coupled Cluster Excitations for Multireference Molecular States Optimized by the Variational Quantum Eigensolver},
  journal = {International Journal of Quantum Chemistry},
  year    = {2021},
  volume  = {121},
  number  = {4},
  pages   = {e26352}
}

@article{ryabinkin2018constrained,
  author  = {Ryabinkin, Ilya G. and Genin, Scott N. and Izmaylov, Artur F.},
  title   = {Constrained Variational Quantum Eigensolver: Quantum Computer Search Engine in the {F}ock Space},
  journal = {Journal of Chemical Theory and Computation},
  year    = {2019},
  volume  = {15},
  number  = {1},
  pages   = {249--255}
}

@article{grimsley2025challenging,
  author  = {Grimsley, Harper R. and Evangelista, Francesco A.},
  title   = {Challenging Excited States from Adaptive Quantum Eigensolvers:
             Subspace Expansions vs. State-Averaged Strategies},
  journal = {Quantum Science and Technology},
  year    = {2025},
  volume  = {10},
  number  = {2},
  pages   = {025003}
}

@misc{sambasivam2025tepid,
  author        = {Sambasivam, Bharath and Sherbert, Kyle and
                   Shirali, Karunya and Mayhall, Nicholas J. and
                   Harrow, Aram W. and Barnes, Edwin and
                   Economou, Sophia E.},
  title         = {{TEPID-ADAPT}: Adaptive Variational Method for
                   Simultaneous Preparation of Low-Temperature Gibbs
                   and Low-Lying Eigenstates},
  year          = {2025},
  archivePrefix = {arXiv},
  primaryClass  = {quant-ph},
  note          = {arXiv:2503.14490v2}
}

@misc{saroni2026tepid,
  author        = {Saroni, Jason and Sambasivam, Bharath and
                   Asthana, Ayush and Barnes, Edwin},
  title         = {Quantum Simulation of Molecular Excited-State
                   Manifolds and Energies Using the
                   {TEPID-ADAPT-VQE} Algorithm},
  year          = {2026},
  archivePrefix = {arXiv},
  primaryClass  = {quant-ph},
  note          = {arXiv:2606.29547v1}
}

@misc{sundar2026molecular,
  author        = {Sundar, Srivathsan Poyyapakkam and
                   Kwao, Prince Frederick and Galda, Alexey and
                   Asthana, Ayush},
  title         = {Molecular Excited States Using Quantum Subspace
                   Methods: Accuracy, Resource Reduction, and
                   Error-Mitigated Hardware Implementation of
                   {q-sc-EOM}},
  year          = {2026},
  archivePrefix = {arXiv},
  primaryClass  = {quant-ph},
  note          = {arXiv:2604.05380v1}
}

@article{parrish2019mcvqe,
  author  = {Parrish, Robert M. and Hohenstein, Edward G. and McMahon, Peter L. and Mart{\'i}nez, Todd J.},
  title   = {Quantum Computation of Electronic Transitions Using a Variational Quantum Eigensolver},
  journal = {Physical Review Letters},
  year    = {2019},
  volume  = {122},
  number  = {23},
  pages   = {230401}
}

@article{kwao2026generalized,
  title={Generalized Eigenvalue Problem in Subspace-Based Excited-State Methods for Quantum Computers},
  author={Kwao, Prince Frederick and Poyyapakkam Sundar, Srivathsan and Gupt, Brajesh and Asthana, Ayush},
  journal={Journal of Chemical Theory and Computation},
  volume={22},
  number={6},
  pages={2892--2903},
  year={2026},
  publisher={ACS Publications}
}

@article{kandala2017hardware,
  author  = {Kandala, Abhinav and Mezzacapo, Antonio and Temme, Kristan and Takita, Maika and Brink, Markus and Chow, Jerry M. and Gambetta, Jay M.},
  title   = {Hardware-efficient variational quantum eigensolver for small molecules and quantum magnets},
  journal = {Nature},
  year    = {2017},
  volume  = {549},
  number  = {7671},
  pages   = {242--246}
}

@article{romero2018strategies,
  author  = {Romero, Jonathan and Babbush, Ryan and McClean, Jarrod R. and Hempel, Cornelius and Love, Peter J. and Aspuru-Guzik, Al{\'a}n},
  title   = {Strategies for quantum computing molecular energies using the unitary coupled cluster ansatz},
  journal = {Quantum Science and Technology},
  year    = {2019},
  volume  = {4},
  number  = {1},
  pages   = {014008}
}

@article{ryabinkin2018qubit,
  author  = {Ryabinkin, Ilya G. and Yen, Tzu-Ching and Genin, Scott N. and Izmaylov, Artur F.},
  title   = {Qubit Coupled Cluster Method: A Systematic Approach to Quantum Chemistry on a Quantum Computer},
  journal = {Journal of Chemical Theory and Computation},
  year    = {2018},
  volume  = {14},
  number  = {12},
  pages   = {6317--6326}
}

@article{grimsley2019adaptive,
  author  = {Grimsley, Harper R. and Economou, Sophia E. and Barnes, Edwin and Mayhall, Nicholas J.},
  title   = {An adaptive variational algorithm for exact molecular simulations on a quantum computer},
  journal = {Nature Communications},
  year    = {2019},
  volume  = {10},
  pages   = {3007}
}

@article{tang2021qubit,
  author  = {Tang, Ho Lun and Shkolnikov, V. O. and Barron, George S. and Grimsley, Harper R. and Mayhall, Nicholas J. and Barnes, Edwin and Economou, Sophia E.},
  title   = {{Qubit-ADAPT-VQE}: An Adaptive Algorithm for Constructing Hardware-Efficient Ans{\"a}tze on a Quantum Processor},
  journal = {PRX Quantum},
  year    = {2021},
  volume  = {2},
  number  = {2},
  pages   = {020310}
}

@article{windom2024ultimate,
  title={An “ultimate” coupled cluster method based entirely on {T$_2$}},
  author={Windom, Zachary W and Perera, Ajith and Bartlett, Rodney J},
  journal={The Journal of Chemical Physics},
  volume={161},
  number={18},
  year={2024},
pages = 184106,
  publisher={AIP Publishing}
}

@article{windom2024attractive,
  title={An attractive way to correct for missing singles excitations in unitary coupled cluster doubles theory},
  author={Windom, Zachary W and Claudino, Daniel and Bartlett, Rodney J},
  journal={The Journal of Physical Chemistry A},
  volume={128},
  number={33},
  pages={7036--7045},
  year={2024},
  publisher={ACS Publications}
}

@incollection{windoma2024t2,
  author    = {Windom, Zachary W. and Bartlett, Rodney J.},
  title     = {An effective,  ``strong''' electron correlation estimate using alternative formulations of {T$_2$}-based coupled cluster theory and the factorization theorem},
  booktitle = {Novel Treatments of Strong Correlations},
  editor    = {Bartlett, Rodney J.},
  publisher = {Elsevier},
  year      = {2024},
  pages     = {1--14}
}

@article{windom2024new,
  title={A new “gold standard”: Perturbative triples corrections in unitary coupled cluster theory and prospects for quantum computing},
  author={Windom, Zachary W and Claudino, Daniel and Bartlett, Rodney J},
  journal={The Journal of Chemical Physics},
  volume={160},
  number={21},
  year={2024},
pages = 214113,
  publisher={AIP Publishing}
}

@article{windom2026analysis,
  title={Analysis of fourth-, fifth-, and infinite-order triple excitations in unitary coupled cluster theory},
  author={Windom, Zachary W and Claudino, Daniel},
  journal={The Journal of Physical Chemistry A},
  volume={130},
  number={7},
  pages={1560--1569},
  year={2026},
  publisher={ACS Publications}
}

@article{nakanishi2019subspace,
  title={Subspace-search variational quantum eigensolver for excited states},
  author={Nakanishi, Ken M and Mitarai, Kosuke and Fujii, Keisuke},
  journal={Physical Review Research},
  volume={1},
  number={3},
  pages={033062},
  year={2019},
  publisher={APS}
}

@article{xie2022orthogonal,
  title={Orthogonal state reduction variational eigensolver for the excited-state calculations on quantum computers},
  author={Xie, Qing-Xing and Liu, Sheng and Zhao, Yan},
  journal={Journal of Chemical Theory and Computation},
  volume={18},
  number={6},
  pages={3737--3746},
  year={2022},
  publisher={ACS Publications}
}

@article{chan2021molecular,
  title={Molecular excited state calculations with adaptive wavefunctions on a quantum eigensolver emulation: reducing circuit depth and separating spin states},
  author={Chan, Hans Hon Sang and Fitzpatrick, Nathan and Segarra-Mart{\'\i}, Javier and Bearpark, Michael J and Tew, David P},
  journal={Physical Chemistry Chemical Physics},
  volume={23},
  number={46},
  pages={26438--26450},
  year={2021},
  publisher={Royal Society of Chemistry}
}

@article{higgott2019variational,
  title={Variational quantum computation of excited states},
  author={Higgott, Oscar and Wang, Daochen and Brierley, Stephen},
  journal={Quantum},
  volume={3},
  pages={156},
  year={2019},
  publisher={Verein zur F{\"o}rderung des Open Access Publizierens in den Quantenwissenschaften}
}

@article{peruzzo2014variational,
  title={A variational eigenvalue solver on a photonic quantum processor},
  author={Peruzzo, Alberto and McClean, Jarrod and Shadbolt, Peter and Yung, Man-Hong and Zhou, Xiao-Qi and Love, Peter J and Aspuru-Guzik, Al{\'a}n and O’Brien, Jeremy L},
  journal={Nature Communications},
  volume={5},
  number={1},
  pages={4213},
  year={2014},
  publisher={Nature Publishing Group UK London}
}

@article{motta2020determining,
  title={Determining eigenstates and thermal states on a quantum computer using quantum imaginary time evolution},
  author={Motta, Mario and Sun, Chong and Tan, Adrian TK and O’Rourke, Matthew J and Ye, Erika and Minnich, Austin J and Brandao, Fernando GSL and Chan, Garnet Kin-Lic},
  journal={Nature Physics},
  volume={16},
  number={2},
  pages={205--210},
  year={2020},
  publisher={Nature Publishing Group UK London}
}

@article{tkachenko2024quantum,
  title={Quantum Davidson algorithm for excited states},
  author={Tkachenko, Nikolay V and Cincio, Lukasz and Boldyrev, Alexander I and Tretiak, Sergei and Dub, Pavel A and Zhang, Yu},
  journal={Quantum Science and Technology},
  volume={9},
  number={3},
  pages={035012},
  year={2024},
  publisher={IOP Publishing}
}

@article{colless2018computation,
  title={Computation of molecular spectra on a quantum processor with an error-resilient algorithm},
  author={Colless, James I and Ramasesh, Vinay V and Dahlen, Dar and Blok, Machiel S and Kimchi-Schwartz, Mollie E and McClean, Jarrod R and Carter, Jonathan and de Jong, Wibe A and Siddiqi, Irfan},
  journal={Physical Review X},
  volume={8},
  number={1},
  pages={011021},
  year={2018},
  publisher={APS}
}

@article{mcclean2020decoding,
  title={Decoding quantum errors with subspace expansions},
  author={McClean, Jarrod R and Jiang, Zhang and Rubin, Nicholas C and Babbush, Ryan and Neven, Hartmut},
  journal={Nature Communications},
  volume={11},
  number={1},
  pages={636},
  year={2020},
  publisher={Nature Publishing Group UK London}
}

@article{takeshita2020increasing,
  title={Increasing the representation accuracy of quantum simulations of chemistry without extra quantum resources},
  author={Takeshita, Tyler and Rubin, Nicholas C and Jiang, Zhang and Lee, Eunseok and Babbush, Ryan and McClean, Jarrod R},
  journal={Physical Review X},
  volume={10},
  number={1},
  pages={011004},
  year={2020},
  publisher={APS}
}

@article{mcclean2017hybrid,
  title={Hybrid quantum-classical hierarchy for mitigation of decoherence and determination of excited states},
  author={McClean, Jarrod R and Kimchi-Schwartz, Mollie E and Carter, Jonathan and De Jong, Wibe A},
  journal={Physical Review A},
  volume={95},
  number={4},
  pages={042308},
  year={2017},
  publisher={APS}
}

@article{kim2018coupled,
  title={Coupled double triplet state in singlet fission},
  author={Kim, Hyungjun and Zimmerman, Paul M},
  journal={Physical Chemistry Chemical Physics},
  volume={20},
  number={48},
  pages={30083--30094},
  year={2018},
  publisher={Royal Society of Chemistry}
}

@article{stanton1993equation,
  title={The equation of motion coupled-cluster method. A systematic biorthogonal approach to molecular excitation energies, transition probabilities, and excited state properties},
  author={Stanton, John F and Bartlett, Rodney J},
  journal={The Journal of Chemical Physics},
  volume={98},
  number={9},
  pages={7029--7039},
  year={1993},
  publisher={AIP Publishing}
}

@article{watson2013benchmarking,
  title={Benchmarking for perturbative triple-excitations in {EE-EOM-CC} methods},
  author={Watson Jr, Thomas J and Lotrich, Victor F and Szalay, Peter G and Perera, Ajith and Bartlett, Rodney J},
  journal={The Journal of Physical Chemistry A},
  volume={117},
  number={12},
  pages={2569--2579},
  year={2013},
  publisher={ACS Publications}
}

@article{rishi2017excited,
  title={Excited states from modified coupled cluster methods: {Are they any better than EOM CCSD?}},
  author={Rishi, Varun and Perera, Ajith and Nooijen, Marcel and Bartlett, Rodney J},
  journal={The Journal of Chemical Physics},
  volume={146},
  number={14},
  year={2017},
  publisher={AIP Publishing}
}

@article{bartlett2007coupled,
  title={Coupled-cluster theory in quantum chemistry},
  author={Bartlett, Rodney J and Musia{\l}, Monika},
  journal={Reviews of Modern Physics.},
  volume={79},
  number={1},
  pages={291},
  year={2007},
  publisher={APS}
}

@article{boguslawski2018targeting,
  title={Targeting doubly excited states with equation of motion coupled cluster theory restricted to double excitations},
  author={Boguslawski, Katharina},
  journal={Journal of Chemical Theory and Computation},
  volume={15},
  number={1},
  pages={18--24},
  year={2018},
  publisher={ACS Publications}
}

@article{kucharski2001coupled,
  title={Coupled-cluster theory for excited electronic states: The full equation-of-motion coupled-cluster single, double, and triple excitation method},
  author={Kucharski, Stanis{\l}aw A and W{\l}och, Marta and Musia{\l}, Monika and Bartlett, Rodney J},
  journal={The Journal of Chemical Physics},
  volume={115},
  number={18},
  pages={8263--8266},
  year={2001},
  publisher={American Institute of Physics}
}

@article{hirata2000high,
  title={High-order determinantal equation-of-motion coupled-cluster calculations for electronic excited states},
  author={Hirata, So and Nooijen, Marcel and Bartlett, Rodney J},
  journal={Chemical Physics Letters},
  volume={326},
  number={3-4},
  pages={255--262},
  year={2000},
  publisher={Elsevier}
}

@article{kallay2001higher,
  title={Higher excitations in coupled-cluster theory},
  author={K{\'a}llay, Mih{\'a}ly and Surj{\'a}n, P{\'e}ter R},
  journal={The Journal of Chemical Physics},
  volume={115},
  number={7},
  pages={2945--2954},
  year={2001},
  publisher={American Institute of Physics}
}

@article{accomasso2022ultrafast,
  title={Ultrafast excited-state dynamics of carotenoids and the role of the Sx state},
  author={Accomasso, Davide and Arslancan, Serra and Cupellini, Lorenzo and Granucci, Giovanni and Mennucci, Benedetta},
  journal={The Journal of Physical Chemistry Letters},
  volume={13},
  number={29},
  pages={6762--6769},
  year={2022},
  publisher={ACS Publications}
}

@article{polivka2004ultrafast,
  title={Ultrafast dynamics of carotenoid excited states- from solution to natural and artificial systems},
  author={Pol{\'\i}vka, Tom{\'a}{\v{s}} and Sundstr{\"o}m, Villy},
  journal={Chemical Reviews},
  volume={104},
  number={4},
  pages={2021--2072},
  year={2004},
  publisher={ACS Publications}
}

@article{mirkovic2017light,
  title={Light absorption and energy transfer in the antenna complexes of photosynthetic organisms},
  author={Mirkovic, Tihana and Ostroumov, Evgeny E and Anna, Jessica M and Van Grondelle, Rienk and Govindjee and Scholes, Gregory D},
  journal={Chemical Reviews},
  volume={117},
  number={2},
  pages={249--293},
  year={2017},
  publisher={ACS Publications}
}

@article{nozik2008multiple,
  title={Multiple exciton generation in semiconductor quantum dots},
  author={Nozik, Arthur J},
  journal={Chemical Physics Letters},
  volume={457},
  number={1-3},
  pages={3--11},
  year={2008},
  publisher={Elsevier}
}

@article{davis2015multiple,
  title={Multiple-exciton generation in lead selenide nanorod solar cells with external quantum efficiencies exceeding 120\%},
  author={Davis, Nathaniel JLK and B{\"o}hm, Marcus L and Tabachnyk, Maxim and Wisnivesky-Rocca-Rivarola, Florencia and Jellicoe, Tom C and Ducati, Caterina and Ehrler, Bruno and Greenham, Neil C},
  journal={Nature Communications},
  volume={6},
  number={1},
  pages={8259},
  year={2015},
  publisher={Nature Publishing Group UK London}
}

@article{scholes2025quantum,
  title={The quantum information science challenge for chemistry},
  author={Scholes, Gregory D and Olaya-Castro, Alexandra and Mukamel, Shaul and Kirrander, Adam and Ni, Kang-Kuen and Hedley, Gordon J and Frank, Natia L},
  journal={The Journal of Physical Chemistry Letters},
  volume={16},
  number={5},
  pages={1376--1396},
  year={2025},
  publisher={ACS Publications}
}

@article{lee2019excited,
  title={Excited states via coupled cluster theory without equation-of-motion methods: Seeking higher roots with application to doubly excited states and double core hole states},
  author={Lee, Joonho and Small, David W and Head-Gordon, Martin},
  journal={The Journal of Chemical Physics},
  volume={151},
  number={21},
  year={2019},
  publisher={AIP Publishing}
}

@article{loos2019reference,
  title={Reference energies for double excitations},
  author={Loos, Pierre-Fran{\c{c}}ois and Boggio-Pasqua, Martial and Scemama, Anthony and Caffarel, Michel and Jacquemin, Denis},
  journal={Journal of Chemical Theory and Computation},
  volume={15},
  number={3},
  pages={1939--1956},
  year={2019},
  publisher={ACS Publications}
}

@article{kossoski2021excited,
  title={Excited states from state-specific orbital-optimized pair coupled cluster},
  author={Kossoski, F{\'a}bris and Marie, Antoine and Scemama, Anthony and Caffarel, Michel and Loos, Pierre-Francois},
  journal={Journal of Chemical Theory and Computation},
  volume={17},
  number={8},
  pages={4756--4768},
  year={2021},
  publisher={ACS Publications}
}

@article{liu2018unitary,
  title={Unitary coupled-cluster based self-consistent polarization propagator theory: A third-order formulation and pilot applications},
  author={Liu, Junzi and Asthana, Ayush and Cheng, Lan and Mukherjee, Debashis},
  journal={The Journal of Chemical Physics},
  volume={148},
  number={24},
  year={2018},
  pages=244110,
  publisher={AIP Publishing}
}

@article{kutzelnigg1983quantum,
  title={Quantum chemistry in {Fock space. II. Effective Hamiltonians in Fock} space},
  author={Kutzelnigg, Werner and Koch, Sigurd},
  journal={The Journal of Chemical Physics},
  volume={79},
  number={9},
  pages={4315--4335},
  year={1983},
  publisher={American Institute of Physics}
}

@article{hodecker2020third,
  title={Third-order unitary coupled cluster (UCC3) for excited electronic states: Efficient implementation and benchmarking},
  author={Hodecker, Manuel and Thielen, Sebastian M and Liu, Junzi and Rehn, Dirk R and Dreuw, Andreas},
  journal={Journal of Chemical Theory and Computation},
  volume={16},
  number={6},
  pages={3654--3663},
  year={2020},
  publisher={ACS Publications}
}

@article{liu2021unitary,
  title={Unitary coupled-cluster based self-consistent polarization propagator theory: A quadratic unitary coupled-cluster singles and doubles scheme},
  author={Liu, Junzi and Cheng, Lan},
  journal={The Journal of Chemical Physics},
  volume={155},
  number={17},
  year={2021},
  pages=174102,
  publisher={AIP Publishing}
}

@article{taube2006new,
  title={New perspectives on unitary coupled-cluster theory},
  author={Taube, Andrew G and Bartlett, Rodney J},
  journal={International Journal of Quantum Chemistry},
  volume={106},
  number={15},
  pages={3393--3401},
  year={2006},
  publisher={Wiley Online Library}
}

@article{bartlett1989alternative,
  title={Alternative coupled-cluster ans{\"a}tze {II. The} unitary coupled-cluster method},
  author={Bartlett, Rodney J and Kucharski, Stanislaw A and Noga, Jozef},
  journal={Chemical Physics Letters},
  volume={155},
  number={1},
  pages={133--140},
  year={1989},
  publisher={Elsevier}
}

@book{shavitt2009many,
  title={Many-Body Methods in Chemistry and Physics: MBPT and Coupled-Cluster Theory},
  author={Shavitt, Isaiah and Bartlett, Rodney J},
  year={2009},
  publisher={Cambridge University Press}
}

@inproceedings{bartlett1989some,
  title={Some consideration of alternative ans{\"a}tz in coupled-cluster theory},
  author={Bartlett, Rodney J and Kucharski, Stanislaw A and Noga, Jozef and Watts, John D and Trucks, Gary W},
  booktitle={Many-Body Methods in Quantum Chemistry: Proceedings of the Symposium, Tel Aviv University 28--30 August 1988},
  pages={125--149},
  year={1989},
  organization={Springer}
}

@article{bartlett1988expectation,
  title={The expectation value coupled-cluster method and analytical energy derivatives},
  author={Bartlett, Rodney J and Noga, Jozef},
  journal={Chemical Physics Letters},
  volume={150},
  number={1-2},
  pages={29--36},
  year={1988},
  publisher={Elsevier}
}

@article{lowdin1962studies,
  title={Studies in perturbation theory. IV. Solution of eigenvalue problem by projection operator formalism},
  author={L{\"o}wdin, Per-Olov},
  journal={Journal of Mathematical Physics},
  volume={3},
  number={5},
  pages={969--982},
  year={1962},
  publisher={American Institute of Physics}
}

@article{kitaev1997quantum,
  title={Quantum computations: algorithms and error correction},
  author={Kitaev, A Yu},
  journal={Russian Mathematical Surveys},
  volume={52},
  number={6},
  pages={1191--1249},
  year={1997}
}

@book{nielsen2010quantum,
  title={Quantum computation and quantum information},
  author={Nielsen, Michael A and Chuang, Isaac L},
  year={2010},
  publisher={Cambridge University Press}
}

@article{russo2021evaluating,
  title={Evaluating energy differences on a quantum computer with robust phase estimation},
  author={Russo, Antonio E and Rudinger, Kenneth Michael and Morrison, Benjamin CA and Baczewski, Andrew David},
  journal={Physical review letters},
  volume={126},
  number={21},
  pages={210501},
  year={2021},
  publisher={APS}
}

@article{bauman2020toward,
  title={Toward quantum computing for high-energy excited states in molecular systems: quantum phase estimations of core-level states},
  author={Bauman, Nicholas P and Liu, Hongbin and Bylaska, Eric J and Krishnamoorthy, Sriram and Low, Guang Hao and Granade, Christopher E and Wiebe, Nathan and Baker, Nathan A and Peng, Bo and Roetteler, Martin and others},
  journal={Journal of Chemical Theory and Computation},
  volume={17},
  number={1},
  pages={201--210},
  year={2020},
  publisher={ACS Publications}
}

@article{sugisaki2021quantum,
  title={Quantum algorithm for the direct calculations of vertical ionization energies},
  author={Sugisaki, Kenji and Toyota, Kazuo and Sato, Kazunobu and Shiomi, Daisuke and Takui, Takeji},
  journal={The Journal of Physical Chemistry Letters},
  volume={12},
  number={11},
  pages={2880--2885},
  year={2021},
  publisher={ACS Publications}
}

@article{santagati2018witnessing,
  title={Witnessing eigenstates for quantum simulation of {H}amiltonian spectra},
  author={Santagati, Raffaele and Wang, Jianwei and Gentile, Antonio A and Paesani, Stefano and Wiebe, Nathan and McClean, Jarrod R and Morley-Short, Sam and Shadbolt, Peter J and Bonneau, Damien and Silverstone, Joshua W and Tew, David P and Zhou, Xiaoqi and O'Brien, Jeremy L and Thompson, Mark G},
  journal={Science Advances},
  volume={4},
  number={1},
  pages={eaap9646},
  year={2018},
  publisher={American Association for the Advancement of Science}
}

@article{o2019quantum,
  title={Quantum phase estimation of multiple eigenvalues for small-scale (noisy) experiments},
  author={O’Brien, Thomas E and Tarasinski, Brian and Terhal, Barbara M},
  journal={New Journal of Physics},
  volume={21},
  number={2},
  pages={023022},
  year={2019},
  publisher={IOP Publishing}
}

@article{parrish2019quantum,
  title={Quantum filter diagonalization: Quantum eigendecomposition without full quantum phase estimation},
  author={Parrish, Robert M and McMahon, Peter L},
  journal={arXiv preprint arXiv:1909.08925},
  year={2019}
}

@article{sugisaki2022adiabatic,
  title={Adiabatic state preparation of correlated wave functions with nonlinear scheduling functions and broken-symmetry wave functions},
  author={Sugisaki, Kenji and Toyota, Kazuo and Sato, Kazunobu and Shiomi, Daisuke and Takui, Takeji},
  journal={Communications Chemistry},
  volume={5},
  number={1},
  pages={84},
  year={2022},
  publisher={Nature Publishing Group UK London}
}

@article{watts1996iterative,
  title={Iterative and non-iterative triple excitation corrections in coupled-cluster methods for excited electronic states: the {EOM-CCSDT-3 and EOM-CCSD (T) methods}},
  author={Watts, John D and Bartlett, Rodney J},
  journal={Chemical Physics Letters},
  volume={258},
  number={5-6},
  pages={581--588},
  year={1996},
  publisher={Elsevier}
}

@article{christiansen1995response,
  title={Response functions in the {CC3} iterative triple excitation model},
  author={Christiansen, Ove and Koch, Henrik and J{\o}rgensen, Poul},
  journal={The Journal of Chemical Physics},
  volume={103},
  number={17},
  pages={7429--7441},
  year={1995},
  publisher={American Institute of Physics}
}

@article{bartlett1990non,
  title={Non-iterative fifth-order triple and quadruple excitation energy corrections in correlated methods},
  author={Bartlett, Rodney J and Watts, JD and Kucharski, SA and Noga, J},
  journal={Chemical Physics Letters},
  volume={165},
  number={6},
  pages={513--522},
  year={1990},
  publisher={Elsevier}
}

@article{watts1995economical,
  title={Economical triple excitation equation-of-motion coupled-cluster methods for excitation energies},
  author={Watts, John D and Bartlett, Rodney J},
  journal={Chemical Physics Letters},
  volume={233},
  number={1-2},
  pages={81--87},
  year={1995},
  publisher={Elsevier}
}

@article{stanton1996simple,
  title={A simple correction to final state energies of doublet radicals described by equation-of-motion coupled cluster theory in the singles and doubles approximation},
  author={Stanton, John F and Gauss, J{\"u}rgen},
  journal={Theoretica Chimica Acta},
  volume={93},
  number={5},
  pages={303--313},
  year={1996},
  publisher={Springer}
}

@article{zhao2024enhancing,
  title={Enhancing the electron pair approximation with measurements on trapped-ion quantum computers},
  author={Zhao, Luning and Wang, Qingfeng and Goings, Joshua J and Shin, Kyujin and Kyoung, Woomin and Noh, Seunghyo and Rhee, Young Min and Kim, Kyungmin},
  journal={npj Quantum Information},
  volume={10},
  number={1},
  pages={76},
  year={2024},
  publisher={Nature Publishing Group UK London}
}

@article{patra2025energy,
  title={Energy landscape plummeting in variational quantum eigensolver: Subspace optimization, non-iterative corrections, and generator-informed initialization for improved quantum efficiency},
  author={Patra, Chayan and Maitra, Rahul},
  journal={The Journal of Chemical Physics},
  volume={163},
  number={2},
  year={2025},
  publisher={AIP Publishing}
}

@article{evangelista2019exact,
  title={Exact parameterization of fermionic wave functions via unitary coupled cluster theory},
  author={Evangelista, Francesco A and Chan, Garnet Kin and Scuseria, Gustavo E},
  journal={The Journal of Chemical Physics},
  volume={151},
  pages={244112},
  number={24},
  year={2019},
  publisher={AIP Publishing}
}

@article{watts1994inclusion,
  title={The inclusion of connected triple excitations in the equation-of-motion coupled-cluster method},
  author={Watts, John D and Bartlett, Rodney J},
  journal={The Journal of Chemical Physics},
  volume={101},
  number={4},
  pages={3073--3078},
  year={1994},
  publisher={American Institute of Physics}
}

@article{koch1990excitation,
  title={{Excitation energies from the coupled cluster singles and doubles linear response function (CCSDLR). Applications to Be, CH$^+$, CO, and H$_2$O}},
  author={Koch, Henrik and Jensen, Hans Jo/rgen Aa and J{\o}rgensen, Poul and Helgaker, Trygve},
  journal={The Journal of Chemical Physics},
  volume={93},
  number={5},
  pages={3345--3350},
  year={1990},
  publisher={American Institute of Physics}
}

@article{geertsen1989equation,
  title={The equation-of-motion coupled-cluster method: Excitation energies of {Be and CO}},
  author={Geertsen, Jan and Rittby, Magnus and Bartlett, Rodney J},
  journal={Chemical Physics Letters},
  volume={164},
  number={1},
  pages={57--62},
  year={1989},
  publisher={Elsevier}
}

@article{ravi2022intermediate,
  title={The intermediate state approach for doubly excited dark states in EOM-coupled-cluster theory},
  author={Ravi, Moneesha and Park, Young Choon and Perera, Ajith and Bartlett, Rodney J},
  journal={The Journal of Chemical Physics},
  volume={156},
  number={20},
  year={2022},
  pages={201102},
  publisher={AIP Publishing}
}

@article{rishi2023dark,
  title={Dark doubly excited states with modified coupled cluster models: A reliable compromise between cost and accuracy?},
  author={Rishi, Varun and Ravi, Moneesha and Perera, Ajith and Bartlett, Rodney J},
  journal={The Journal of Physical Chemistry A},
  volume={127},
  number={3},
  pages={828--834},
  year={2023},
  publisher={ACS Publications}
}

@article{ravi2023excited,
  title={Excited states with pair coupled cluster doubles tailored coupled cluster theory},
  author={Ravi, Moneesha and Perera, Ajith and Park, Young Choon and Bartlett, Rodney J},
  journal={The Journal of Chemical Physics},
  volume={159},
  number={9},
  year={2023},
  publisher={AIP Publishing}
}

@article{d2023challenges,
  title={Challenges in the use of quantum computing hardware-efficient ansätze in electronic structure theory},
  author={D’Cunha, Ruhee and Crawford, T Daniel and Motta, Mario and Rice, Julia E},
  journal={The Journal of Physical Chemistry A},
  volume={127},
  number={15},
  pages={3437--3448},
  year={2023},
  publisher={ACS Publications}
}

@article{Freericks2022,
  title = {Operator Relationship between Conventional Coupled Cluster and Unitary Coupled Cluster},
  volume = {14},
  number = {3},
  journal = {Symmetry},
  publisher = {MDPI AG},
  author = {Freericks,  James K.},
  year = {2022},
  month = feb,
  pages = {494}
}

@article{zheng2024quantum,
  title={Quantum equation-of-motion method with single, double, and triple excitations},
  author={Zheng, Yuhan and Sun, Zhijie and Liu, Jie and Fan, Yi and Li, Zhenyu and Yang, Jinlong},
  journal={Journal of Chemical Theory and Computation},
  volume={20},
  number={20},
  pages={9032--9040},
  year={2024},
  publisher={ACS Publications}
}

@article{Motta2023,
  title = {Bridging physical intuition and hardware efficiency for correlated electronic states: the local unitary cluster Jastrow ansatz for electronic structure},
  volume = {14},
  ISSN = {2041-6539},
  number = {40},
  journal = {Chemical Science},
  publisher = {Royal Society of Chemistry (RSC)},
  author = {Motta,  Mario and Sung,  Kevin J. and Whaley,  K. Birgitta and Head-Gordon,  Martin and Shee,  James},
  year = {2023},
  pages = {11213–11227}
}

\end{document}